\documentclass[fleqn,usenatbib]{mnras}

\usepackage{newtxtext,newtxmath}

\usepackage[T1]{fontenc}

\DeclareRobustCommand{\VAN}[3]{#2}
\let\VANthebibliography\thebibliography
\def\thebibliography{\DeclareRobustCommand{\VAN}[3]{##3}\VANthebibliography}

\usepackage{graphicx}	
\usepackage{amsmath}	
\usepackage{caption}
\usepackage{subcaption}

\title[PG BL Lacs with uGMRT]{A Kpc-scale Radio Polarization Study of PG BL Lacs with the uGMRT}

\author[J. Baghel et al.]{
Janhavi Baghel,$^{1}$\thanks{E-mail: jbaghel@ncra.tifr.res.in}
P. Kharb,$^{1}$
T. Hovatta,$^{2,3}$
S. Gulati,$^{1,4}$
E. Lindfors,$^{2}$
Silpa S.$^{1}$
\\
$^{1}$National Centre for Radio Astrophysics (NCRA) - Tata Institute of Fundamental Research (TIFR), S. P.  Pune University Campus, Ganeshkhind, Pune 411007, India \\
$^{2}$Finnish Centre for Astronomy with ESO, FINCA, University of Turku, Turku, Finland\\
$^{3}$Aalto University Metsähovi Radio Observatory, Metsähovintie 114, 02540 Kylmälä, Finland\\
$^{4}$Manipal Centre for Natural Sciences, Centre of Excellence, Manipal Academy of Higher Education, Manipal 576104, India\\\\
}

\begin{document}
\label{firstpage}
\pagerange{\pageref{firstpage}--\pageref{lastpage}}
\maketitle

\begin{abstract}
We present here uGMRT band 4 ($\sim$650~MHz) polarization images of 8 BL~Lac objects belonging to the Palomar-Green (PG) `blazar' sample. A large fraction of the sources ($\sim63\%$) reveal core-halo radio structures with most of the polarization detected in the inner core-jet regions. PG1101+385 and PG2254+075 exhibit a `spine-sheath structure' in polarization. The core-halo and `spine-sheath' structures are consistent with the Unified Scheme suggestion that BL~Lacs are the pole-on beamed counterparts of Fanaroff-Riley (FR) type I radio galaxies. PG1418+546 and PG0851+203 (OJ287) show the presence of terminal hotspots similar to FR type II radio galaxies. They were also found to be low-spectrally peaked BL Lacs, supportive of the `blazar envelope' scenario for BL~Lacs and quasars. Fractional polarization ranges from $1-13$\% in the cores and $2-26$\% in the inner jets/lobes of the sample BL~Lacs. Compared to the varied radio morphology of quasars from the PG `blazar' sample, 
the BL~Lacs appear to be less diverse. A comparison of the inferred core magnetic (B-) field structures on arcsec- (kpc-) scales w.r.t. the Very Long Baseline Interferometry (VLBI) jet direction does not reveal any preferred orientation, suggesting that if large-scale ordered B-fields exist, they do so on scales smaller than probed by the current observations. However, the presence of polarized emission on arcsec-scales suggests that any mixing of thermal plasma with the synchrotron emitting plasma is insufficient to fully depolarize the emission via the internal depolarization process. 
\end{abstract}

\section{Introduction}
BL~Lac objects are a sub-class of highly variable and energetic active galactic nuclei (AGN) called `blazars', characterized by their high luminosities, rapid variability, high and variable polarization, superluminal jet motion, and intense non-thermal emission across the entire electromagnetic spectrum \citep{Angel1980,Urry95}. Blazars are `radio-loud' (RL) sources: the ratio of their radio 5~GHz to optical B-band flux densities is, $\mathrm{R\equiv S_{5~GHz}/S_{B-band}} \geq 10$ \citep{Kellermann1989}. The extreme properties of blazars are now understood to be largely due to their relativistic jets being oriented at small angles to our lines of sight, where both Doppler-boosting and projection effects become prominent \citep{Blandford1978, Blandford1979}. 

BL~Lacs are suggested to be the pole-on counterparts of the Fanaroff-Riley type I (FRI) radio galaxies \citep{Fanaroff1974}, while RL quasars are the pole-on counterparts of FRII radio galaxies under the radio-loud unification scheme \citep{Urry95}. Quasars were originally distinguished from BL~Lacs by the presence of broad (equivalent widths of $> 5$ \AA ) optical emission lines, the BL~Lac spectra being mostly featureless and dominated by continuum emission \citep{Stickel1991,Stocke1991}. The spectral energy distributions (SEDs) of both subclasses of blazars show two peaks, with the first ascribed to synchrotron emission. Based on this synchrotron peak frequency ($\nu_s$), blazars are classified into low synchrotron peaked objects (LSP; $\nu_s < 10^{14}$ Hz), intermediate synchrotron peaked objects (ISP; $\nu_s = 10^{14} - 10^{15}$ Hz), and high synchrotron peaked objects (HSP; $\nu_s > 10^{15}$ Hz). BL Lac objects occupy all these three classes whereas quasars are only found to be LSPs \citep{Abdo2010}. 

It is noteworthy that the simple radio-loud unification scheme fails for several blazar samples that are not radio-selected. For example, the MOJAVE\footnote{Monitoring Of Jets in Active galactic nuclei with VLBA Experiments, \citet{Lister2018}.} \citep[e.g.,][]{Kharb10}, which is selected based on the total parsec-scale radio luminosity at 15~GHz, or the Palomar-Green (PG) sample \citep{Green1986} that is UV-excess selected \citep[e.g.,][]{Baghel2023}. Several BL~Lac objects either exhibit FRII-like radio powers or have morphologies like FRIIs (i.e., terminal hotspots), and FRI-like morphology is also seen in quasars \citep{Landt2006,Kharb10}. 

It was \citet{Fossati1998} who noted a systematic anti-correlation of radio luminosity with $\nu_s$ and introduced the notion of a `blazar sequence'. This was later revised with Fermi data by \citet{Ghisellini2017} who found that BL~Lacs and quasars show different behaviors in their SEDs. The previous trends existed for BL Lac objects, while for FSRQs $\nu_s$ remained constant with luminosity. In BL Lacs, the two SED peaks had almost equal luminosities, while in FSRQs the second peak was significantly higher, i.e., the SED was Compton dominated. An increase in the luminosity in FSRQs only increased the amount of Compton dominance \citep{Ghisellini2017}. These were interpreted as differences in the radiative cooling mechanisms of the two blazar classes \citep{Ghisellini1998,Ghisellini2013}. 

Jet formation mechanisms, either through Blandford-Znajek \citep[BZ;][]{BZ1977} or Blandford-Payne \citep[BP;][]{BP1982}, or a combination of both, may also be driving the FR or blazar divide. Several theoretical models of AGN jets predict the generation of helical B-fields propagating outward with the jet plasma \citep{Meier2001, Lyutikov2005, Hawley2015}. Polarization observations are therefore important to understand and differentiate between the different jet formation mechanisms, as well as to probe the nature of the magneto-ionised media surrounding these jets. In terms of polarization studies of the blazar parent population of FR radio galaxies, studies indicate that at low radio frequencies, fractional polarization up to 40\% is common in radio jets \citep{Bridle1984}. The inferred B-field structures are mostly parallel to the jet axis in the more powerful FRII sources, whereas for FRI sources B-fields can be mostly perpendicular to the jet axis, or perpendicular to the jet axis at the jet's centre but parallel near one or both of its edges \citep{Willis1981,Bridle1982,Laing2014}. The estimated B-field direction in the jets of most strong quasars after correcting for Faraday rotation is generally found to be along the jets, often closely following jet bends \citep{Bridle1994}. 

In terms of the radio spectral index, it was found by \citet{Laing2013} that the spectrum of FRI sources flattens as one moves away from the AGN along the jet. This was interpreted in terms of a shift in the parameters of the jet's ongoing particle acceleration, either by first-order Fermi process in mildly relativistic shocks present throughout the jet volume or by multiple modes of acceleration being present in the jet, one with an inner high-speed `spine' of the jet and the other associated with the slower speed `sheath' of jet material surrounding it \citep{Laing2013, Laing2014}. The latter is also consistent with the `spine-sheath structure' of polarization often found in these jets at both parsec and kpc scales \citep{Pushkarev2005, Kharb2008, Laing2014}.

Very long baseline interferometry (VLBI) observations of blazars have found differences among the two blazar classes in the magnetic (B-) field structures as well as rotation measures (RM) on parsec-scales \citep{Gabuzda1992, Cawthrone1993}. BL~Lacs tended to have their parsec-scale electric vector position angles (EVPA) parallel to the jet direction whereas RL quasars tended to show a perpendicular relative orientation \citep{Lister2005, Lister2013}. The inferred B-field structures are perpendicular to the EVPAs for optically thin emission, and parallel to the EVPAs for optically thick emission\footnote{The optical thickness required to change the inference of B-field direction is considered to be $\tau \sim 7$ {\citep{Cawthrone2013, Wardle2018}}.} \citep{Pacholczyk70}. 

Quasars show systematically higher parsec-scale RM compared to the BL~Lacs \citep{Zavala2004, Hovatta2012}. For both BL Lacs and quasars, \citet{Pushkarev2017} discovered a considerable rise in the degree of linear polarization with distance from the radio core along the jet and also towards the jet borders. This rise in polarization with distance might be due to an increase in B-field ordering as we proceed down the jet. The increase in polarization around jet edges is ascribed to a fall in depolarization there than along the jet axis. The BL Lacs also have more polarised cores and jets, with the EVPA in both cores and jets seeming to be more aligned with the jet axis (i.e., B-fields perpendicular to the jet direction) compared to quasars which showed orthogonal orientation \citep[i.e., B-fields parallel to the jet direction;][]{Pushkarev2023}.

However, it is not clear if these differences extend to kpc scales. Compared to quasars, BL Lacs have not been as well studied at radio frequencies, on several different spatial scales \citep{Giroletti2006}. Detailed VLBI studies on BL~Lac objects have been carried out only on the brightest of sources like from the 1 Jy sample \citep{Cassaro2002}. The spatially extended diffuse and unbeamed emission is potentially observable at low radio frequencies \citep{Hardcastle2014, Massaglia2016}. \citet{Aller1999} found in their University of Michigan Radio Astronomy Observatory (UMRAO) BL~Lac sample that changes in polarized flux density and EVPAs are common in BL~Lacs and the long-term preferred position angle (PA; measured from North through East with North being at $0\degr$) orientation is uncorrelated to the VLBI jet flow direction. \citet{Fan2006} have studied the radio polarization of BL~Lac objects at 5, 8, and 15~GHz using UMRAO and found that: (1) polarization at higher radio frequencies is greater than at lower frequencies, (2) polarization variability at higher frequencies is greater than variability at lower frequencies, (3) polarization is correlated with the radio spectral index, and (4) polarization is correlated with the core-dominance parameter, implying that relativistic beaming could explain the polarization characteristic of BL~Lacs.

In this paper, we present polarization-sensitive upgraded Giant Metrewave Radio Telescope (uGMRT) images at 650~MHz of 8 BL~Lac objects belonging to the PG `blazar' sample. Here we have adopted the $\Lambda$CDM cosmology with $\mathrm{H_0 = 73~km~s^{-1} Mpc^{-1}}$, $\mathrm{\Omega_m = 0.3}$ and $\mathrm{\Omega_v = 0.7}$ 
and used flat $\Lambda$CDM subroutine of astropy.cosmology subpackage \citep{astropy:2013,astropy:2018}. The spectral index $\alpha$ is defined such that flux density at frequency $\nu$, $S_\nu \propto \nu^\alpha$.

\begin{table*}
 \label{tab1}
\centering
\caption{BL Lacs from the PG `blazar' sample.
}
\begin{tabular}{c|c|c|c|c|c|c|c|c}
\hline
S.No. & Name  & Other Name  & RA & DEC & Redshift & Radio Extent ($\arcsec$ | kpc) & AGN Class & R\\
\hline
1 & {PG0851+203} & 
OJ287 & 
08h54m48.87s &
+20{\degr}06{\arcmin}30.64{\arcsec} &
0.3056 &
20 | 87 &
LSP & 
1,235.68
\\
2 &
{PG1101+385} &
Mrk421 &
11h04m27.31s &
+38{\degr}12{\arcmin}31.79{\arcsec} &
0.03002 &
30 | 18  &
HSP & 
34.94
\\
3 &
{PG1218+304} &
RBS 1100 &
12h21m21.94s &
+30{\degr}10{\arcmin}37.16{\arcsec} &
0.1836 &
45 | 134 &
HSP & 
27.74
\\
4 &
{PG1418+546} &
OQ +530 &
14h19m46.59s &
+54{\degr}23{\arcmin}14.78{\arcsec} &
0.15248 &
45 | 115 &
LSP & 
620.86
\\
5 &
{PG1424+240} &
OQ +240&
14h27m00.39s &
+23{\degr}48{\arcmin}00.03{\arcsec} &
0.604$^\ast$ &
45 | 290 &
HSP & 
134.04
\\
6 &
{PG1437+398} &
RBS 1414 &
14h39m17.47s &
+39{\degr}32{\arcmin}42.80{\arcsec} &
0.34366 &
25 | 117 &
HSP & 
34.30
\\
7 &
{PG1553+113} &
RBS 1538 &
15h55m43.04s &
+11{\degr}11{\arcmin}24.36{\arcsec} &
0.36 &
60 | 290 &
HSP & 
120.83
\\
8 &
{PG 2254+075} &
OY +091 &
22h57m17.30s &
+07{\degr}43{\arcmin}12.30{\arcsec} &
0.19 &
60 | 181 &
LSP &
649.32
\\
\hline
\multicolumn{9}{l}{Note. Column (1): Serial Number. Column (2): PG names of sources. Column (3): Other common names of sources. Column (4): Right Ascension.}\\
\multicolumn{9}{l}{Column (5): Declination. Column (6): Redshift$^\dagger$. Column (7): Radio extents in arcsec and kpc. Column (9): SED Class of BL Lac object } \\
\multicolumn{9}{l}{from \citet{Ajello2022}. Column (10): Radio loudness parameter.}\\
\multicolumn{9}{l}{$^\ast$ $z=0.604$ as suggested by \citet{Paiano17}. It was earlier reported as 0.16 in literature and hence included in the sample.}\\
\multicolumn{9}{l}{The redshift of this object is uncertain. Current limits place it in the range z = 0.24 - 1.19 \citep[e.g.][]{Rovero2016,Zahoor2022} }\\
\multicolumn{9}{l}{Radio extents derived from \citet{Miller1993} at 5~GHz or from 1.4~GHz VLA FIRST / NVSS images for  sources unresolved at 5~GHz.}\\
\multicolumn{9}{l}{Radio loudness parameter R derived using B-band magnitudes from \citet{Schmidt1983} and 5~GHz flux densities from NASA NED.}\\
\multicolumn{9}{l}{$^\dagger$ All redshift values reported are corrected to the reference frame defined by the 3K CMB on NASA NED
\footnote{The NASA/IPAC Extragalactic Database (NED) is operated by the Jet Propulsion Laboratory, California Institute of Technology, under contract with the National Aeronautics and Space Administration.
} }
\end{tabular}
\end{table*}

\subsection{The PG `blazar' Sample}
The complete PG sample with a limiting B-band magnitude of 16.1 mag contains 1715 objects and covers a 10,714 square degrees region of the sky. Extragalactic sources comprise only 9\% of the PG sample with quasi-stellar objects (QSOs) constituting 5.4\% of the sample. These QSOs include both radio-quiet quasars ($>80$\%) and radio-loud quasars. Only four BL~Lac objects were initially identified in the PG sample: PG0851+203 (OJ287), PG1418+546, PG1553+113, and PG1218+304. \citet{Padovani1995} revised this list to nine: apart from the original four, three BL~Lacs were identified by \citet{Fleming1993} (earlier misclassified as white dwarfs), and PG1101+384 (Mrk421) and PG2254+075 by \citet{Padovani1995}. \citet{Padovani1995} have noted that the PG sample may be missing 40\% of BL~Lacs because of its U-B color limit. 

The PG `blazar' sample as described by \citet{Baghel2023} comprises 16 RL quasars and 8 BL~Lac objects. This sample was chosen based on a redshift cutoff ($z<0.5$) as well as core-to-lobe extents being greater than $\ge 15 \arcsec$ so that they could be studied with enough resolution with the Karl G. Jansky Very Large Array (VLA) and uGMRT. \citet{Baghel2023} presented polarimetric images at 5~GHz from the VLA of 9 RL quasars and found that a large fraction of them showed distorted or hybrid radio structures and possible restarted jet activity. In the present paper, we present polarization-sensitive uGMRT images at 650~MHz of the 8 BL~Lac objects\footnote{The BL Lac PG1246+586 with a redshift of 0.8 was excluded from the PG `blazar' sample} from the PG `blazar' sample. 

\section{Radio Data Analysis}
Radio data were acquired with the uGMRT in Band-4 (650~MHz) from 2022, January 21 to 2022, May 14 (Project ID: 42\_091, DDTC217 and DDTC273) with a resolution of $\sim5\arcsec$. 2048 channels were chosen to span the frequency range of 550–850~MHz. The average time on source was around $\sim 2$~hr for project 42\_091, $\sim1$~hr for DDTC217, and $\sim 2$~hr for DDTC273. Polarization calibrators 3C286 and 3C138 were used. For more details, see Table~\ref{tab2}. The data were reduced using the \texttt{CASA}\footnote{Common Astronomy Software Applications; \citet{CASA2022}} based polarization pipeline for uGMRT \footnote{https://github.com/jbaghel/Improved-uGMRT-polarization-pipeline} \citep[see also][]{Silpa2021}. The polarization images were created using AIPS\footnote{Astronomical Image Processing System \citep{vanMoorsel1996}}.

The pipeline involves converting the uGMRT data retrieved in the Long-Term Archive (LTA) format to Flexible Image Transport System (FITS) format and then to Measurement Set (MS) file. The data is then flagged and a round of basic gain calibration is done which includes the phase, delay, bandpass, and complex gain calibrations. After applying the calibrations on the MS file additional flagging is done, previously applied calibrations are cleared and another round of basic calibration is performed to get better calibration solutions. These basic calibration steps are followed by polarization calibration.

\begin{table*}
 \caption{Observation Details}
 \label{tab2}
 \begin{tabular}{|c|c|c|c|c|c|}
  \hline
 Source & Observation Date & Flux Calibrator & Phase Calibrator & Leakage Calibrator & Polarization Angle Calibrator \\
 \hline 
 {PG0851+203} & {2022 Jan 21} & 3C286 & J0842+185 & 3C286 & 3C286 \\
 {PG1101+385} & {2022 May 6} & 3C147,3C286 & J1146+399& 3C147 & 3C286\\
 {PG1218+304} & {2022 May 7} & 3C147,3C286 & J1221+282 & 3C147 & 3C286\\ 
 {PG1418+546} & {2022 May 9} & 3C147,3C286 & J1400+621 & 3C147 & 3C286\\
 {PG1424+240} & {2022 Feb 8} & 3C286 & 3C287 & 3C286 & 3C286 \\
 {PG1437+398} & {2022 May 14} & 3C147,3C286 & J1506+375 & 3C147 & 3C286 \\
 {PG1553+113} & {2022 Feb 15} & 3C286 & 3C327.1 & 3C286 & 3C286  \\
 {PG2254+075} & {2022 May 8} & 3C147 & J2241+098 & 3C147 & 3C138\\
  \hline
  \multicolumn{6}{l}{Note. Column(1): PG source name. Column(2): Observation date. Column(3): Flux calibrator. Column(4): Phase calibrator. }\\
  \multicolumn{6}{l}{Column(5): Leakage calibrator. Column(6): Polarization angle calibrator}
 \end{tabular}
\end{table*}

For polarization calibration, a model of the polarization angle calibrator containing its polarization angle, polarization fraction, flux density, and spectral index at the reference frequency needs to be set. For this, the pipeline contains a module for computing these models at the reference frequency of the data using the \citet{Perley2017} scale. The values and errors of the polarization angle and polarization fraction of the polarization angle calibrator are found by fitting a third-order frequency-dependent polynomial. Similarly, the values of spectral index and curvature were found by fitting a second-order frequency-dependent polynomial. These steps were based on the example given in \href{https://casaguides.nrao.edu/index.php?title=CASA_Guides:Polarization_Calibration_based_on_CASA_pipeline_standard_reduction:_The_radio_galaxy_3C75-CASA6.2.1}{CASA Guides}.

Once the model is set, polarization calibration is carried out in the following three steps:
 (1) Cross-hand Delay Calibration: the cross-hand (RL, LR) delays, were solved using a polarized calibrator with strong cross-polarization (either 3C138 or 3C286). This was carried out using the task \texttt{GAINCAL} with gaintype = \texttt{KCROSS} in CASA.
 (2) Leakage Calibration: the instrumental polarization (i.e., the frequency-dependent leakage terms or ‘D-terms’), was solved using a polarized calibrator with good parallactic angle coverage. We used the task \texttt{POLCAL} in CASA to solve for instrumental polarization with poltype = Df + QU while using the polarized calibrators (3C286) and poltype = Df while using the unpolarized calibrator (3C147). The average value of the D-term amplitude turned out to be $\approx$ 20\%.
 (3) Polarization Angle Calibration: the frequency-dependent polarization angle was solved using a polarized calibrator with a known EVPA (either 3C138 or 3C286). This was carried out using the task \texttt{POLCAL} in CASA with poltype = Xf.

The final calibration tables were then applied to the MS file. Additional flagging was done on the target followed by splitting of the calibrated target field data from the MS file with spectral channels averaged taking bandwidth smearing into account. The task \texttt{TCLEAN} was used to obtain the Stokes I, Q, U images of the target source. The multiterm-multifrequency synthesis \citep[MT-MFS;][]{MTMFS2011} algorithm of the \texttt{TCLEAN} task was used. Subsequently, self-calibration cycles were run and the Stokes I image, Stokes Q, and Stokes U images were made from the self-calibrated visibilities. 

After obtaining the images, \texttt{AIPS} was used to get the linear polarized intensity and polarization angle images. The linear polarization intensity were obtained as $\sqrt{\mathrm{Q}^2+\mathrm{U}^2}$ and the polarization fraction as $(\sqrt{\mathrm{Q}^2+\mathrm{U}^2})/{\mathrm{I}}$. The polarization angle ($\chi$) was obtained as $\frac{1}{2}\rm{\tan^{-1}}{\left(\frac{\mathrm{U}}{\mathrm{Q}}\right)}$. In \texttt{AIPS}, the polarization intensity image was made with the task \texttt{comb} and opcode \texttt{POLC} on the Q and U images, clipping below the signal-to-noise ratio (SNR) of 3. The polarization angle image was made with \texttt{COMB} using opcode \texttt{POLA} on the Q and U images, clipping above a noise (polarization angle error) of 10 degrees. To get the fractional polarization image, we used \texttt{COMB} with opcode \texttt{DIV} on the polarization intensity and I images, clipping above a noise (polarization fraction error) of 10\%.  

The spectral index images were made using archival VLA images at 1.4~GHz and our uGMRT images by first producing uGMRT images identical in cellsize, image size convolved with the same beam-size and then aligned with \texttt{AIPS} task \texttt{OGEOM}. AIPS task COMB was then used with opcode `SPIX' to create the spectral index images with pixels below 3 sigma flagged. We also blanked the pixels with spectral index errors greater than 0.3. Flux density values reported here were obtained using the Gaussian-fitting AIPS task \texttt{JMFIT} for compact components like the core, and AIPS verb \texttt{TVSTAT} for extended emission. The fractional polarization (FP) and spectral index values noted are the mean values over the specified regions. The r.m.s. noise values were obtained using AIPS tasks \texttt{TVWIN} and \texttt{IMSTAT}. The AIPS procedure \texttt{TVDIST} was used to obtain all spatial extents.

We note that we have made no attempt to correct for ionospheric Faraday rotation. The principal reason for this being the lack of reliable ionospheric models for the uGMRT. \citet{Farnes2014} had found the maximum Faraday rotation for their 610~MHz legacy GMRT observations due to the ionosphere to be $\le2$~rad~m$^{-2}$. They had used Global Positioning System (GPS) measurements of the vertical total electron content (TEC) from stations in nearby cities and assumed the magnetic field to be weakly variable over the course of their observations. Ionospheric corrections of the order of $0.1-0.3$~rad~m$^{-2}$ have been noted for LOFAR observations at 150~MHz by \citet{Mahatma2021}. We have therefore assumed the ionospheric Faraday rotation effects to not be significant during our observations.

\section{Results}
The radio morphologies of the BL~Lac objects are typically core-halo-like with dominant cores, making them consistent within the unification scheme to be FRI radio galaxies viewed at small inclination angles. Two sources, viz., PG0851+203 and PG1418+546, show the presence of a single hotspot apart from the core with a faint jet or lobe connecting the two. In only one BL Lac object, PG1101+385, a clear two-sided FRI-lobe-like structure is observed. Polarization is detected in the cores and inner jets/lobes of all 8 BL~Lac objects. The fractional polarization ranges from $1-13$\% in the cores and $2-26$\% in the inner jets/lobes. 

We have looked at the mean jet PA from VLBI measurements \citep[][]{Lister2021} and compared the core EVPAs with respect to this parsec-scale jet direction, ahead. We find that for a majority of our sources, the core spectral index is relatively flat but not overly inverted\footnote{i.e., the optical density $\tau=7$ is not reached {\citep[e.g.,][]{Wardle2018}}}. We, therefore, regard the uGMRT cores to be practically optically thin, for the purpose of inferring the B-field orientations. For optically thin emission, the inferred B-fields are perpendicular to the EVPAs. We discuss the inferred B-fields and other properties for individual sources below.

We have looked at various global correlations for the PG `blazar' sample using the most recent and reliable estimates of black hole masses, accretion rates, and SFR values. As far as possible, we have relied on estimates obtained uniformly for all the BL Lac objects to remove errors due to different estimation methods being used. These properties are listed in Table~\ref{tab5}. We have used the Kendall-$\tau$ test to measure the significance of correlations between different properties. Figures \ref{corr1}-\ref{corr3} display the various correlations examined, which are discussed ahead.

\subsection{Notes on individual sources}

\begin{figure*}
\centering
\begin{subfigure}{.5\linewidth}
  \centering
  \includegraphics[width=.6\linewidth, trim= 150 250 150 250]{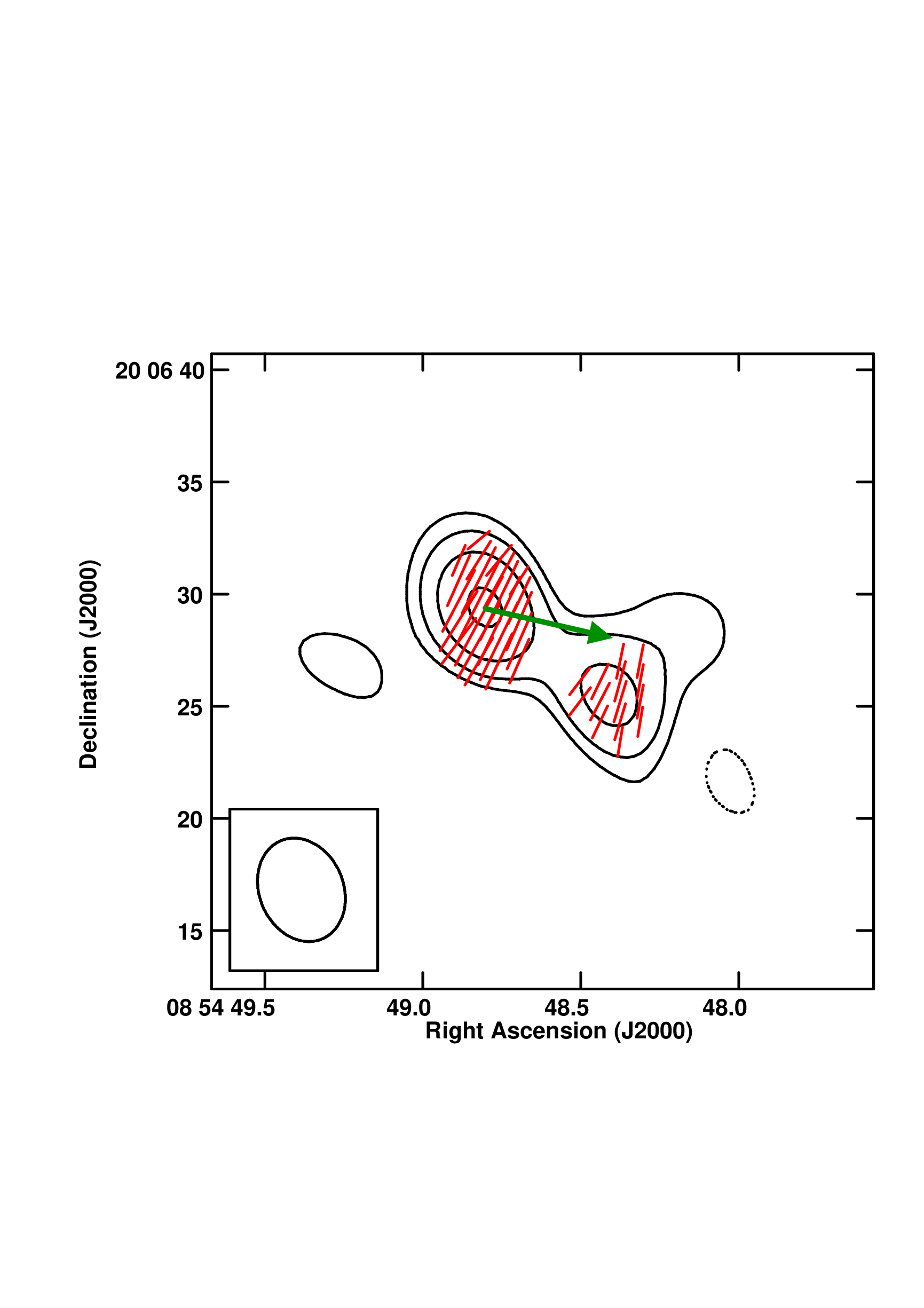}
  
\end{subfigure}%
\begin{subfigure}{.5\linewidth}
  \centering
  \includegraphics[width=.6\linewidth, trim= 60 60 60 60]{OJ287_alpha2_clr.eps}
  
\end{subfigure}
\caption{(Left) uGMRT contour image of the BL Lac object PG0851+203 at 650~MHz with global RM corrected EVPA vectors superimposed as red ticks and VLBI jet direction shown by the green arrow. The length of the ticks is proportional to polarized intensity: $2 \arcsec$ length corresponds to $3.33$~mJy~beam$^{-1}$. The contour levels are $1.029 \times 10^{-2} \times (\pm 11.25, 22.5, 45, 90)$~Jy~beam$^{-1}$. {The beam is $4.75 \times 3.75$ at a PA=$22.93\degr$.}
(Right) 650~MHz uGMRT - 1.45~GHz VLA spectral index image in colour superimposed by contours of uGMRT image convolved with the VLA beam. The contour levels are $1.32 \times 10^{-2} \times (\pm 5.6, 11.25, 22.5, 45, 90)$~Jy~beam$^{-1}$. {The beam is $12.70 \times 11.18$ at a PA=$-79.41\degr$.}}
\label{fig1:OJ287}
\end{figure*}

\begin{figure*}  
\centering{\includegraphics[width=13.2cm, trim = 150 250 100 290]{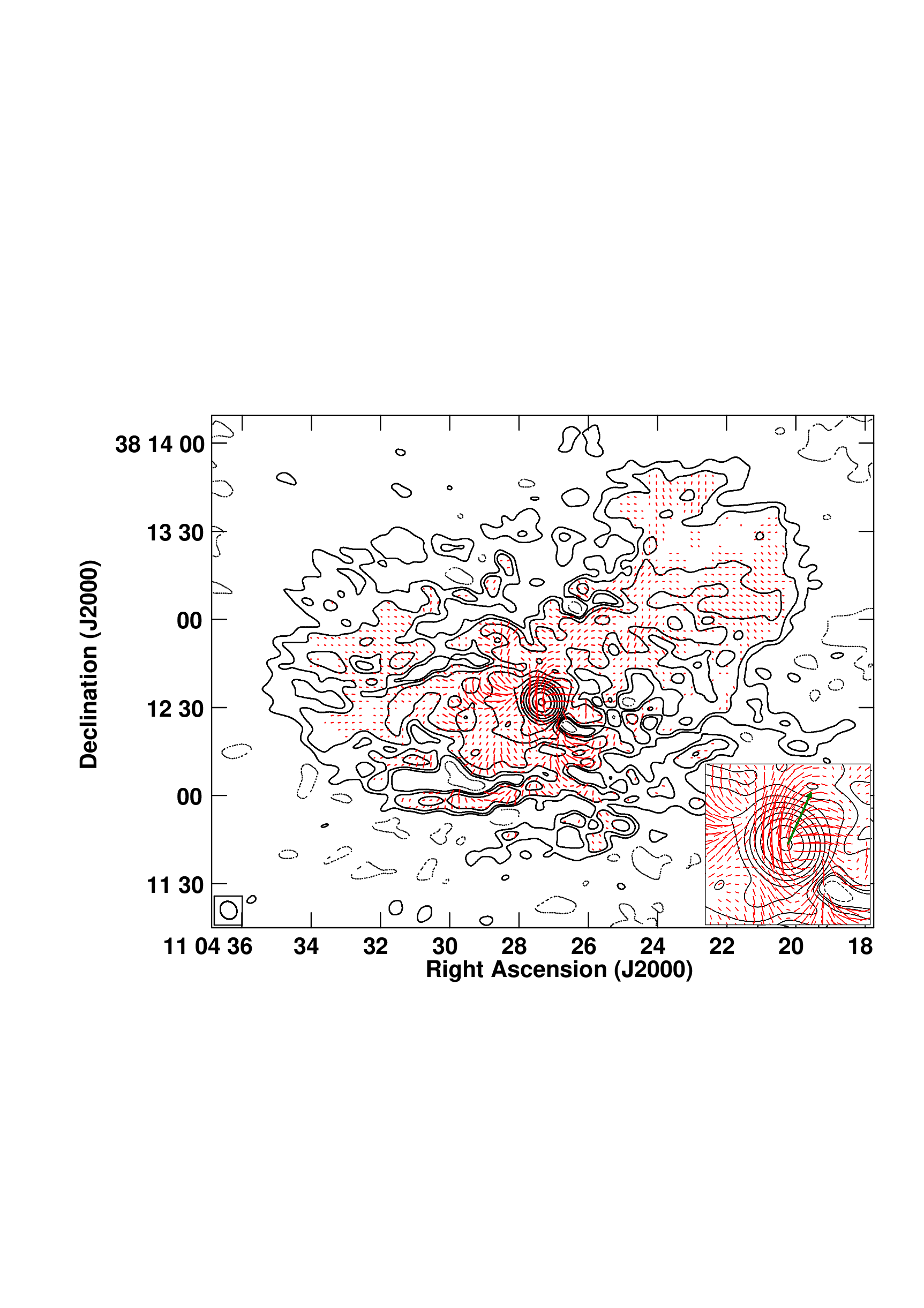} 
\includegraphics[width=8.2cm, trim= 0 60 0 60]{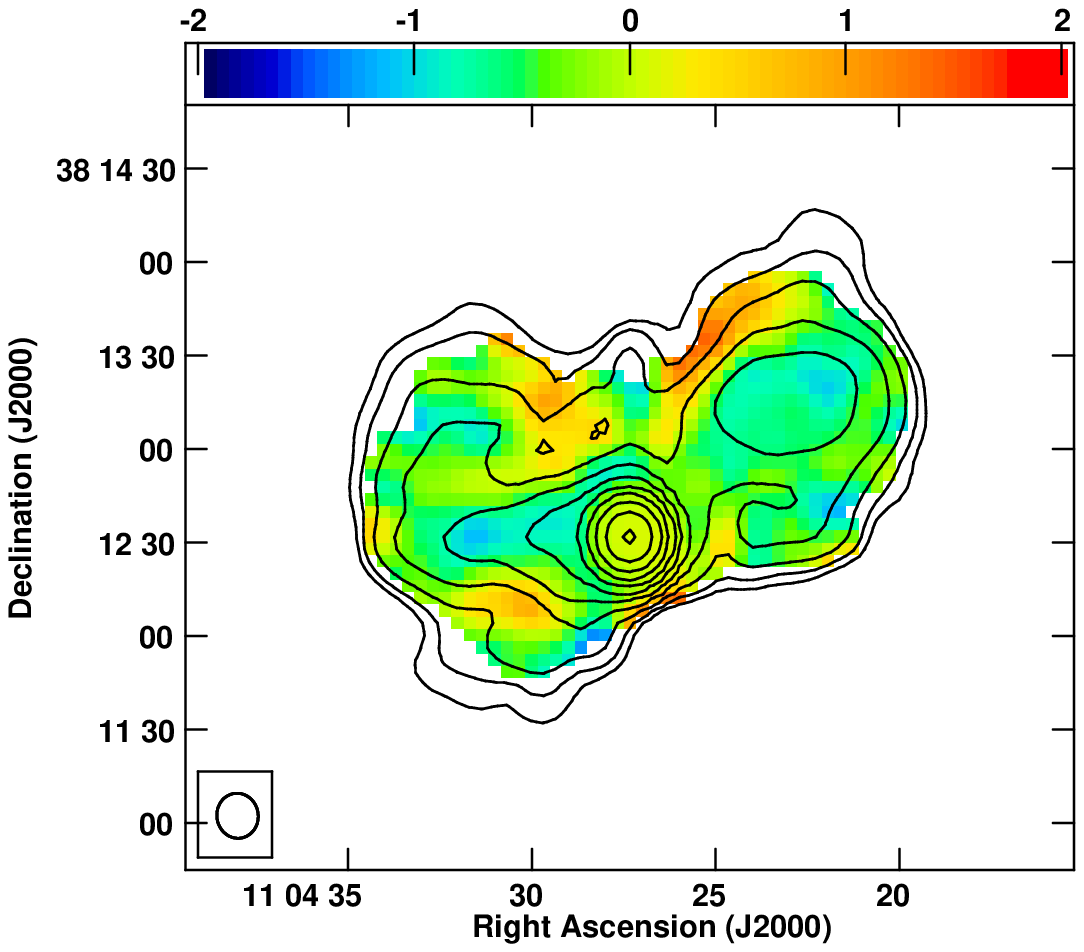}}
\caption{(Left) uGMRT contour image of the BL Lac object PG1101+385 at 650~MHz with global RM corrected EVPA vectors superimposed as red ticks. The inset contains a zoomed-in image of the core region with VLBI jet direction shown by the green arrow. The length of the ticks is proportional to polarized intensity: $20 \arcsec$ length corresponds to $1.33\times10^{-2}$~Jy~beam$^{-1}$. The contour levels are $5.2 \times 10^{-3} \times (\pm 0.09, 0.18, 0.35, 0.70, 1.4, 2.8, 5.6, 11.25, 22.5, 45, 90)$~Jy~beam$^{-1}$. {The beam is $6.27 \times 5.51$ at a PA=$32.49\degr$.} (Right) 650~MHz uGMRT - 1.45~GHz VLA spectral index image in colour superimposed by uGMRT contours where the image is convolved with VLA beam. The contour levels are $5.6 \times 10^{-3} \times (0.09, 0.18, 0.35, 0.70, 1.4, 2.8, 5.6, 11.25, 22.5, 45, 90)$~Jy~beam$^{-1}$. {The beam is $14.47 \times 13.26$ at a PA=$8.28\degr$.} }
\label{fig2:Mkn421}
\end{figure*}

\subsubsection{PG0851+203 a.k.a. OJ287} OJ287 is one of the most well-studied blazars from radio to X-ray frequencies. It is an LSP blazar \citep{Lister2018,Ajello2022}. Its optical light curve which has been extensively studied for well over a century, shows a periodicity of about 12 years \citep{Sillanpaa1988,Sillanpaa1996,Shi2007}. These double-peaked periodic outbursts have been suggested to come from a binary black hole system \citep{Sillanpaa1988,Valtonen2006}. The Very Energetic Radiation Imaging Telescope Array System (VERITAS) has detected very high energy (VHE) emission (>100 GeV) from OJ287 with the observed flux density around 3\% of the Crab Nebula flux in VHE \citep{Mukherjee2017}. 

OJ287 exhibits rapid flux and polarization variations in both radio and optical frequencies. The radio flux changes on a daily timescale at 6.6~GHz and higher frequencies \citep{Kinman1975,Agudo2012}. With VLBI observations at 5~GHz, \citet{1989Gabuzda,Gabuzda1992} found that the source exhibits superluminal motion and that the jet is oriented at a PA of $-100\degr$, aligned with the VLA kpc-scale jet. However, the polarization vectors at milliarcsecond (mas) scales are not well aligned with the VLBI jet. \citet{Myserlis2018} have studied the VLBI polarization data at 15 and 43~GHz and discovered that the EVPA showed a substantial clockwise rotation by $\sim$340$\degr$ with a mean rate of $-1.04\degr$/day. \citet{Lister2021} have provided a mean jet PA with 15~GHz VLBA observations as $-103\degr$. \citet{Gomez2022} used RadioAstron at 22~GHz to get polarimetric VLBI observations of OJ287. The EVPA results showed that the innermost jet has a primarily perpendicular (toroidal) magnetic field. This, along with the marginal evidence of a rotation measure gradient across the width of the jet, suggests that the VLBI core has a helical magnetic field. The stacked 15~GHz VLBA image shows EVPAs perpendicular to the jet direction in the core \citep{Pushkarev2023}.

At radio frequencies, OJ287 possesses a weak tail, first detected by \citet{deBruyn1986} in their 5 GHz Westerbork Synthesis Radio Telescope (WSRT) observations. In the VLA 1.4~GHz A-array ($\sim 1\arcsec$) observations by \citet{Perlman1994}, a jet that is $25\arcsec$ in length with a bend at $15\arcsec$ from the core is seen. There is also a weak terminal hotspot. \citet{Marscher2011} have provided a crude 1.4 GHz polarization image of OJ287, where the r.m.s. noise in the polarized intensity is too high to estimate a polarization angle. A recent multi-wavelength overview of OJ287 jet from parsec- to kpc-scales has been given by \citet{Butuzova2021}.

Our uGMRT 650~MHz image (Figure \ref{fig1:OJ287}) shows a polarized core along with a polarized jet-like component to the southwest. We find the inferred B-fields to be parallel in the core with respect to the VLBI jet direction and also parallel in the kpc-scale jet in a jet knot about $10\arcsec$ away from the core. While we see the beginning of the jet curvature seen clearly in the VLA image of \citet{Perlman1994} we do not detect the terminal hotspot in our uGMRT image. The spectral index image created using archival VLA data (beam-size $12.70\arcsec \times 11.18\arcsec$) shows a flat core and a steepening towards the direction of the jet although the jet itself is not clearly visible because of the poor resolution of the images used for the spectral index image.

\begin{table*}
 \caption{Polarization Properties}
 \label{tab3}
 \begin{tabular}{|c|c|c|c|c|c|c|c|c|c|}
  \hline
Source & $I_\mathrm{rms}$(mJy~beam$^{-1}$) & $P_\mathrm{rms}$(mJy~beam$^{-1}$) & Region && P~(mJy)& I~(Jy) & FP~(\%) & $\alpha$ & $R_c$ \\
 \hline 
 {PG0851+203} & 12.84 & 0.57 & Core & & $7\pm1$  & 0.992 & $1.6\pm0.8$ & $-0.33\pm0.02$ & 3.80\\
 & & & Jet & & $1.8\pm0.5$  & 0.54 & $\le1.0^\ast$ & $-1.7\pm0.2$\\
 {PG1101+385} & 0.17 & 0.04 & Core & & $11.3\pm0.3$ & 0.55 & $5.1\pm0.2$ & $-0.14\pm0.02$ & 3.05\\
 & & & Lobe & & $180\pm20$ & 0.52 & $26\pm6$ & $-0.41 \pm 0.09$ \\
 {PG1218+304} & 0.26 & 0.07 & Core & & $1.7\pm0.2$ & 0.05 & $3.3\pm0.7$ & $0.18 \pm 0.06$ & 12.87\\ 
 {PG1418+546} & 0.28 & 1.77 & Core & & $41\pm4$ & 0.85 & $13\pm2$ & $-0.35\pm0.01$ & 4.43\\
 & & & Hotspot & & $3.2\pm0.8$ & 0.045 & $18\pm5$ & $-0.21\pm 0.07$\\
 {PG1424+240}  & 3.69 & 0.30 & Core & & $3.6\pm0.3$ & 0.24 & $\le1.0^\ast$ & $0.4\pm0.1$  & 0.021\\
  &  &  & N Jet & & $5.9\pm0.9$  & 0.12 & $3.5\pm0.3$  & $-1.1\pm0.2$ \\
  &  &  & S Jet & & $3.1\pm0.5$  & 0.12 & $2.1\pm0.4$ & $-0.6\pm0.2$ \\
 {PG1437+398} & 0.06 & 0.02 & Core & & $1.84\pm0.09$ & 0.0552 & $6.0\pm0.9$ & $-0.24\pm0.08$ & 0.79\\
 {PG1553+113} & 0.65 & 0.02 & Core & & $ 0.94\pm 0.05$& 0.0874 & $\le1.0^\ast$ & $0.23\pm 0.06$ & 1.26
 \\
 {PG2254+075} & 0.33 & 0.11 & Core & & $28.8\pm0.4$ & 0.316 & $9.6\pm0.7$ &  $0.21\pm0.01$ & 0.47\\
  \hline
  \multicolumn{10}{l}{Note. Column (1): PG source name. Column (2): R.M.S noise in Stokes I (total intensity) image. Column (3):  R.M.S noise in the polarized intensity image. }\\
  \multicolumn{10}{l}{Column (4): Region of the source. Column (5): Polarized flux density. Column (6): Total flux density. Column (7): Fractional Polarization. } \\
\multicolumn{10}{l}{Column (8): Spectral index.  Column (9):  Radio Core prominence. }\\
\multicolumn{10}{l}{$\ast$ This is the polarization detection threshold for uGMRT as discussed in \citet{Kharb23}.}
 \end{tabular}
\end{table*}

\begin{table*}
 \caption{Global RM and VLBI Data}
 \label{tab4}
 \begin{tabular}{|c|c|c|c|c|c|c|}
  \hline
  Source & Global RM (rad~m$^{-2}$) & EVPA Rotation at 650~MHz & VLBI jet PA & Ref & EVPA core (15~GHz VLBA) & EVPA core (650~MHz uGMRT)\\ 
 \hline 
 {PG0851+203} & 31.2 $\pm$ 0.5 & $380\degr$ & $-103\degr$ & 1 & $\perp$ & $\perp$\\
 {PG1101+385} & -3.8 $\pm$ 4.5 & $-46\degr$ &  $-23\degr$ & 1 & oblique/ $\parallel$ & $\parallel$ \\
 {PG1218+304} & ... & ... & $+90\degr$ & 3 & ... & $\perp$\\ 
 {PG1418+546} & 18.8 $\pm$ 3.6 & $229\degr$ & $+130\degr$ & 1 & oblique & $\parallel$ \\
 {PG1424+240} & 31.5 $\pm$ 16.4 & $384\degr$ & $+144\degr$ & 1 & $\parallel$ & $\perp$ \\
 {PG1437+398} & ... & ... &... & ... & ... & ...\\
 {PG1553+113} & 27.1 $\pm$ 14.3 & $330\degr$ & $+85\degr$ & 1 & $\parallel$ & oblique/$\perp$ \\
 {PG2254+075} & 0.2 $\pm$ 3.4 & $2\degr$ & $-135\degr$ & 1 & $\parallel$/oblique & $\parallel$\\
  \hline
  \multicolumn{7}{l}{Note. Column (1): PG Source names. Column (2): Global rotation measure from \citet{Taylor2009}. Column (3): EVPA rotation due to RM at 650~MHz.}\\
 \multicolumn{7}{l}{Column (4): Mean VLBI jet position angle. Column (5): References for mean VLBI jet position angle.}\\
  \multicolumn{7}{l}{Column (6): EVPA in core vs VLBI jet (15~GHz VLBA stacked) \citep{Pushkarev2023}. Column (7): EVPA in core vs VLBI jet (650~MHz uGMRT)}\\ 
   \multicolumn{7}{l}{References: (1) \citet{Lister2021} (3) \citet{Giroletti2004}}\\
 \end{tabular}
\end{table*}

\subsubsection{PG1101+385 a.k.a. Mrk421} Mrk421 is one of the brightest $\gamma$-ray sources in the sky and the closest known BL Lac object. It shows strong and rapid TeV outbursts \citep{Punch1992} and is an HSP blazar \citep{Lister2018,Ajello2022}. It is a highly variable source with it flux above 400 GeV varying by more than an order of magnitude in observations taken with MAGIC {(Major Atmospheric Gamma Imaging Cherenkov)} between March 2007 and June 2009 \citep{Ahnen2016}. Its recent Imaging X-Ray Polarimetry Explorer (IXPE) observations of X-ray polarization \citep{Laura2022} show $15\pm2\%$ steady polarization, much higher than that at optical and lower wavelengths suggesting the particles in the jet have a large range of energies. The X-ray EVPA of $+35\degr \pm 4\degr$, is close to the optical value of $+21\degr \pm 1\degr$, but different from the jet direction of $-14\degr \pm 14\degr$ measured at 43 GHz suggesting jet bending between at the sites of high- and low-frequency emission \citep{Marscher2008}. 

\citet{Xu1995} have provided a 5~GHz tapered VLBI image, with a resolution of $11\times8$ mas; revealing a compact core with a complex $0.1\arcsec$ jet at a PA of $\sim -60\degr$. \citet{Zhang1990,Zhang1991} present 5 and 22~GHz VLBI images. A compact core with a 7 mas jet at a PA of $\sim -45\degr$ is seen on the 1 mas resolution 5~GHz image while only an unresolved core can be seen with the 22~GHz image of 0.15 mas resolution. Additionally, they feature a Multi-Element Radio Linked Interferometer Network (MERLIN) image at 5~GHz with a resolution of 65 mas, which displays knots in the jet upto $2.2\arcsec$ away from the core. The European VLBI Network (EVN) and MERLIN images presented by \citet{Giroletti2006} reveal that the incoming jet is approaching from the north-west with a viewing angle of $\leq 20\degr$. No superluminal motion has been detected in the parsec-scale jet of Mrk421 \citep{Piner2010}; the maximum apparent speed found in the MOJAVE survey is $\sim$0.3c. At 22~GHz, the VLBA polarization image from 2006 displays a `spine-sheath structure' in polarization, with EVPAs parallel to the jet at the center and perpendicular to the jet along the edges \citep{Piner2010}. \citet{Jorstad2017} provide 43~GHz VLBA images which show its parsec-scale structure dominated by quasi-stationary features. Limb brightening is also seen in these images. \citet{Lister2019} note that all three of the innermost jet features of Mrk421 seen in their 15~GHz VLBA observations show inward motion. \citet{Lister2021} have provided a mean jet PA with 15~GHz VLBA observations as $-23\degr$. The stacked 15~GHz VLBA image shows EVPA perpendicular to the jet direction in the core and parallel thereafter \citep{Pushkarev2023}.

Initial evidence for a sizeable halo around Mrk421 was found using 2.7 and 8.1~GHz observations from the Green Bank three-element interferometer \citep{Margon1978}. The 1.4 GHz VLA C-array image with $15\arcsec$ resolution by \citet{Machalski1985} reveals it to be a bent triple radio source, with a compact core surrounded by a faint lobe with an angular size upto $3\arcmin$ at a PA of $\sim -60\degr$ and a variable core flux. The Cambridge Low-Frequency Synthesis Telescope (CLFST) also detects and resolves the source at 151~MHz \citep{Minns2000}. The bright compact core is likely to be the source of variability observed at these low frequencies. The core is also variable at higher frequencies \citep[e.g.][{at $\sim10$~GHz}]{Seielstad1983}. \citet{Mooney2021} present the extended morphology of this source at 144~MHz with LOFAR (Low-Frequency Array), which matches the morphology seen in the 1.4~GHz VLA C-array image. \citet{Laurent1993} have provided a low-resolution polarization image of this source with the VLA C-configuration at 1.5~GHz. Their image shows magnetic fields that are highly organized and oriented along the circumference of the lobes' outer edges as often seen for FRII sources \citep{Saikia1988,Kharb2008b}. 

Our 650~MHz uGMRT image (Figure \ref{fig2:Mkn421}) shows a polarized core along with a polarized halo matching the previously observed highly ordered magnetic fields in the lobes. We pick up the signatures of aligned B-fields at the lobe edges and perpendicular B-fields in the central regions of the outflow reminiscent of a kpc-scale `spine-sheath' structure previously observed in the VLA image of Mrk421 by \citet{Laurent1993}. We find the inferred B-fields to be perpendicular to the VLBI jet direction in the core. The spectral index image shows a flat core and spectral steepening towards the lobe.

\begin{figure*}  
\centering
\includegraphics[width=9.4cm, trim = 50 205 50 205]{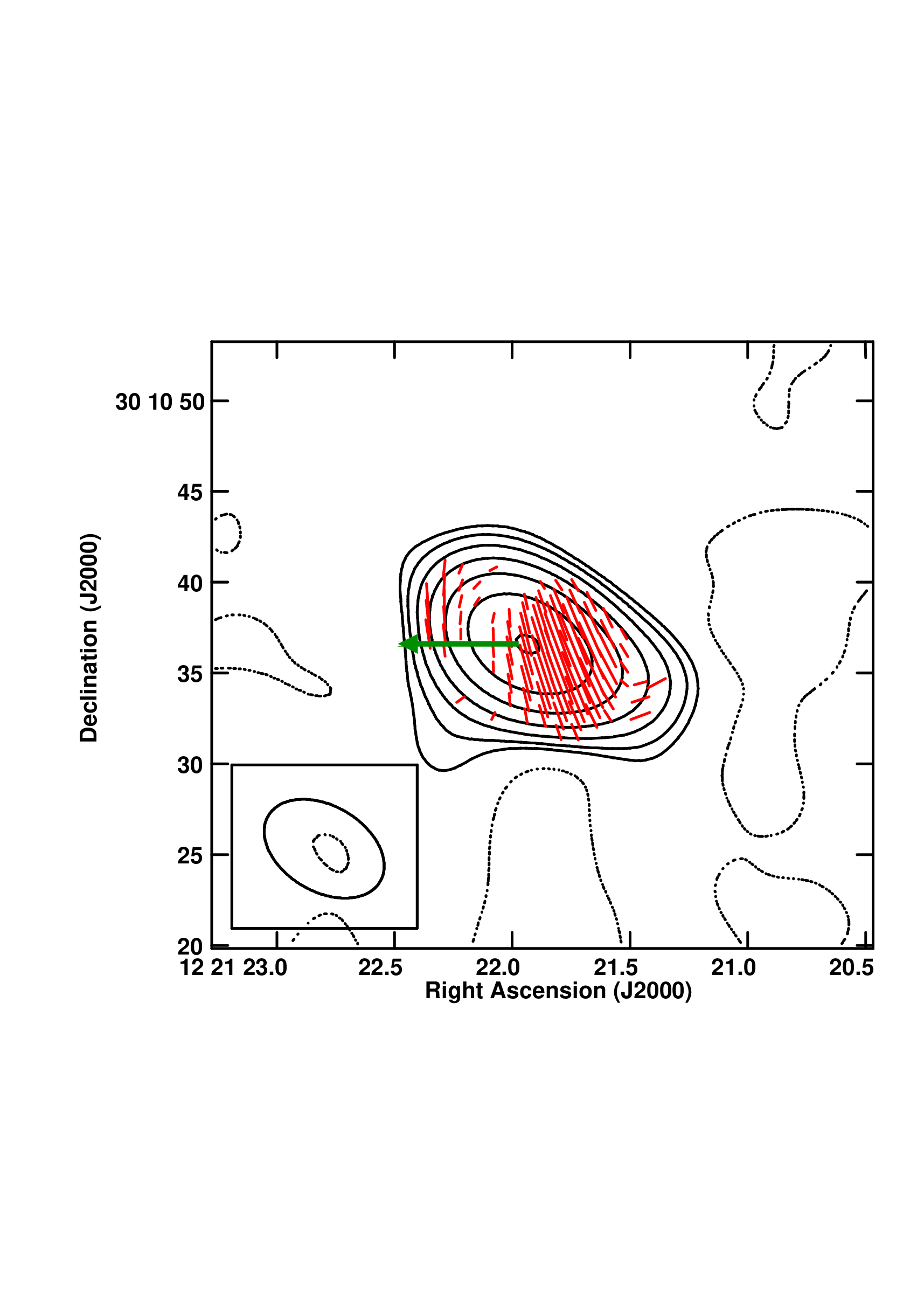}
\includegraphics[width=8.3cm]{PG1218+304_SPIX2.eps}
\caption{(Left) uGMRT contour image of the BL Lac object PG1218+304 at 650~MHz with EVPA vectors superimposed as red ticks and VLBI jet direction shown by the green arrow. The length of the ticks are proportional to polarized intensity. The length of the ticks is proportional to polarized intensity: $5 \arcsec$ length corresponds to $2.77\times10^{-3}$~Jy~beam$^{-1}$. The contour levels are $2.46 \times 10^{-3} \times (\pm 0.35, 0.70, 1.4, 2.8, 5.6, 11.25, 22.5, 45, 90)$~Jy~beam$^{-1}$. {The beam is $7.14 \times 4.72$ at a PA=59.61$\degr$.} (Right) 650~MHz uGMRT - 1.45~GHz VLA spectral index image in colour superimposed by uGMRT contours where the image is convolved with VLA beam. The contour levels are $1.89 \times 10^{-3} \times (\pm 1.4, 2.8, 5.6, 11.25, 22.5)$~Jy~beam$^{-1}$. {The beam is $13.36 \times 10.88$ at a PA=$37.75\degr$.}}
\label{fig4:PG1218}
\end{figure*}

\begin{figure*}  
\centering
\includegraphics[width=9.4cm, trim = 50 300 50 300]{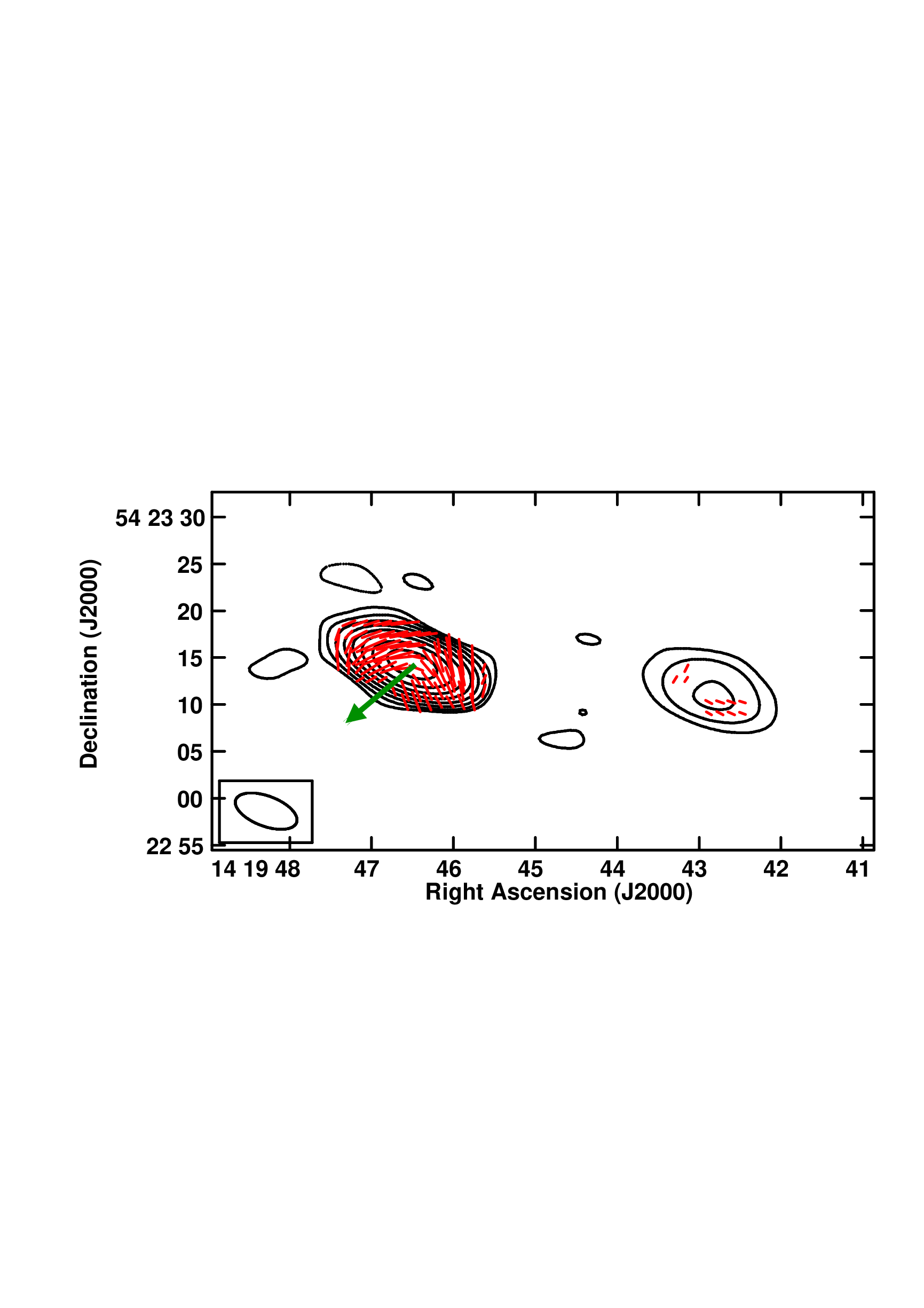}
\includegraphics[width=8.2cm, trim = 0 50 0 0]{PG1418+546_SPIX.eps}
\caption{(Left) uGMRT contour image of the BL Lac object PG1418+546 at 650~MHz with global RM corrected EVPA vectors superimposed as red ticks and VLBI jet direction shown by the green arrow. The lengths of the ticks are proportional to polarized intensity. The length of the ticks is proportional to polarized intensity: $5 \arcsec$ length corresponds to $13.9$~mJy~beam$^{-1}$. The contour levels are $8.48 \times 10^{-3} \times (\pm 0.5, 1, 2, 4, 8, 16, 32, 64)$~Jy~beam$^{-1}$. {The beam is $6.93 \times 3.22$ at a PA=$69.42\degr$.} (Right) 650~MHz uGMRT - 1.45~GHz VLA spectral index image in colour superimposed by uGMRT contours where the image is convolved with VLA beam. The contour levels are $8.67 \times 10^{-3} \times (0.35, 0.7, 1.4, 2.8, 5.6, 11.25, 22.5)$~Jy~beam$^{-1}$. {The beam is $15.16 \times 9.60$ at a PA=$68.20\degr$.}}
\label{fig5:PG1418}
\end{figure*}

\subsubsection{PG1218+304} After its discovery as an unidentified X-ray source in the 2A catalog \citep{Cooke1978}, this source was identified as a BL~Lac object using radio and optical observations in 1979 \citep{Wilson1979}. It has been extensively monitored at optical frequencies and has been found to be highly variable \citep{Jannuzi1993}. It is an HSP blazar \citep{Lister2018,Ajello2022}. The first VHE $\gamma$-ray signal from PG1218+304 with a 6.4$\sigma$ significance above an energy threshold of $\sim$0.12 TeV was discovered by the MAGIC telescope in 2005 \citep{Albert2006}. The RoboPol blazar monitoring program and the Nordic Optical Telescope (NOT) have found an average polarization percentage of about $4.6\%$ and an average optical EVPA of $65\degr$ \citep{Hovatta2016}.

Its VLBA image at 15.4~GHz shows a jet extension to the east \citep{Lister2018} also seen as a 10 mas jet in the 5~GHz VLBA image by \citet{Giroletti2004} (PA $\sim 90\degr$). The source's overall flux density is dominated by the 57 mJy total VLBA correlated flux density \citep{Giroletti2004}. The faint radio images of this blazar at 20 cm were produced by the FIRST {(Faint Images of the Radio Sky at Twenty-Centimeters)} survey using the observations from 1993 April–May \citep{Becker1995}. 

We have found this source to be compact on kpc scales consistent with previous observations \citep[e.g.][and references within]{Perlman1994, Laurent1993}. Our uGMRT 650~MHz image (Figure \ref{fig4:PG1218}) also shows a polarized compact core that has a steep spectrum. The inferred B-field is parallel (poloidal) to the VLBI jet axis.

\begin{figure*}
\centering
\begin{subfigure}{.45\linewidth}
  \centering
  \includegraphics[width=.5\linewidth, trim= 200 250 200 250]{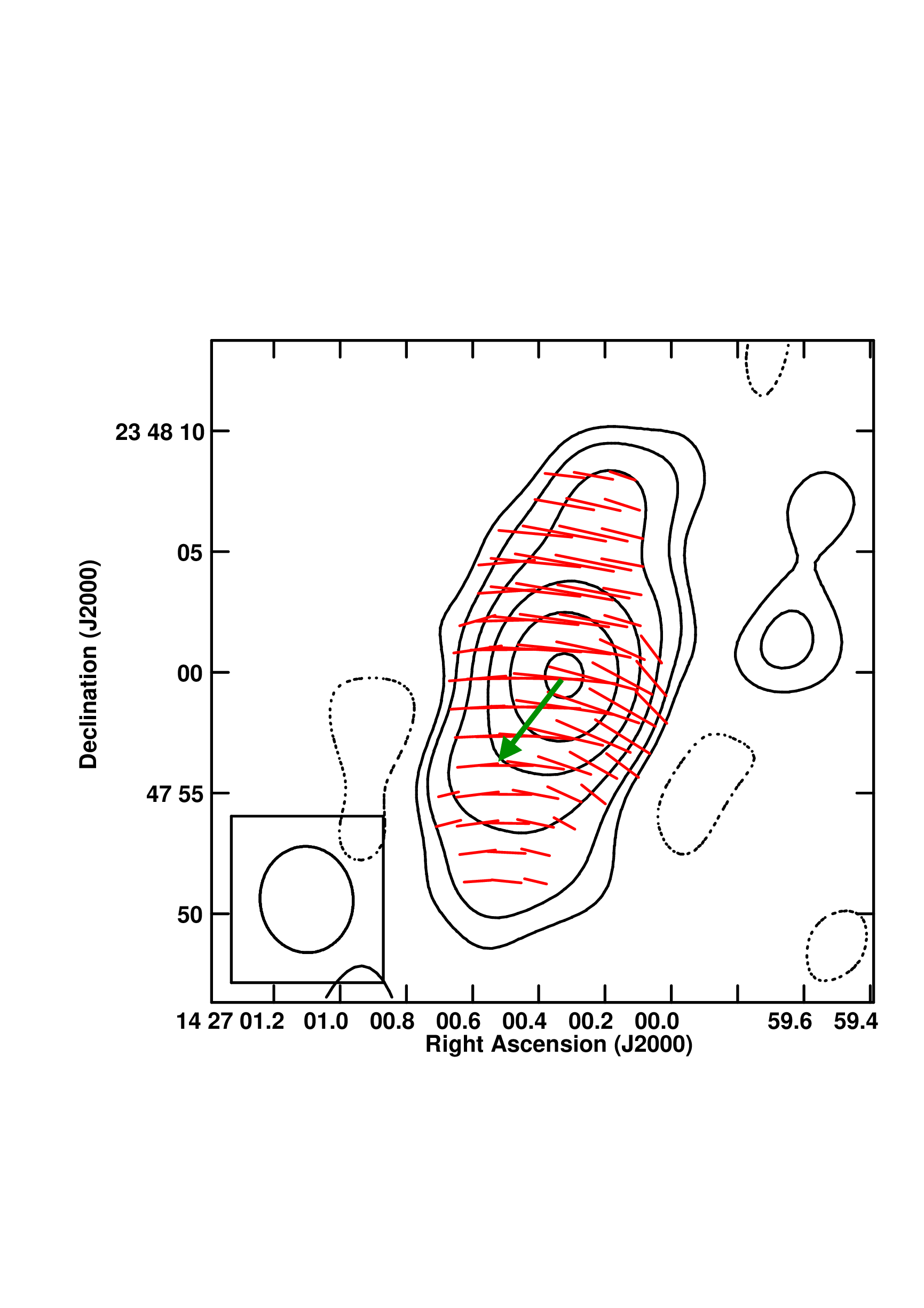}
\end{subfigure}
\begin{subfigure}{.45\linewidth}
  \centering
  \includegraphics[width=.5\linewidth, trim= 100 70 100 70]{PG1424_new_alpha_clr_zm.eps} 
\end{subfigure}

\caption{(Left) uGMRT contour image of the BL Lac object PG1424+240 at 650~MHz with global RM corrected EVPA vectors superimposed as red ticks and VLBI jet direction shown by the green arrow. The length of the ticks is proportional to polarized intensity:  $2 \arcsec$ length corresponds to $1.67$~mJy~beam$^{-1}$. The contour levels are $3.087 \times 10^{-3} \times (\pm 2.8, 5.6, 11.25, 22.5 45, 90)$~Jy~beam$^{-1}$. {The beam is $4.16 \times 3.60$ at a PA=$13.90\degr$.} 
(Right) 650~MHz uGMRT - 1.45~GHz VLA spectral index image in colour superimposed by contours of uGMRT image convolved with the VLA beam.  The contour levels are $4.493 \times 10^{-3} \times (\pm 5.6, 11.25, 22.5, 45, 90)$~Jy~beam$^{-1}$. {The beam is $17.64 \times 14.29$ at a PA=$75.12\degr$.} }
\label{fig6:PG1424}
\end{figure*}

\begin{figure*}
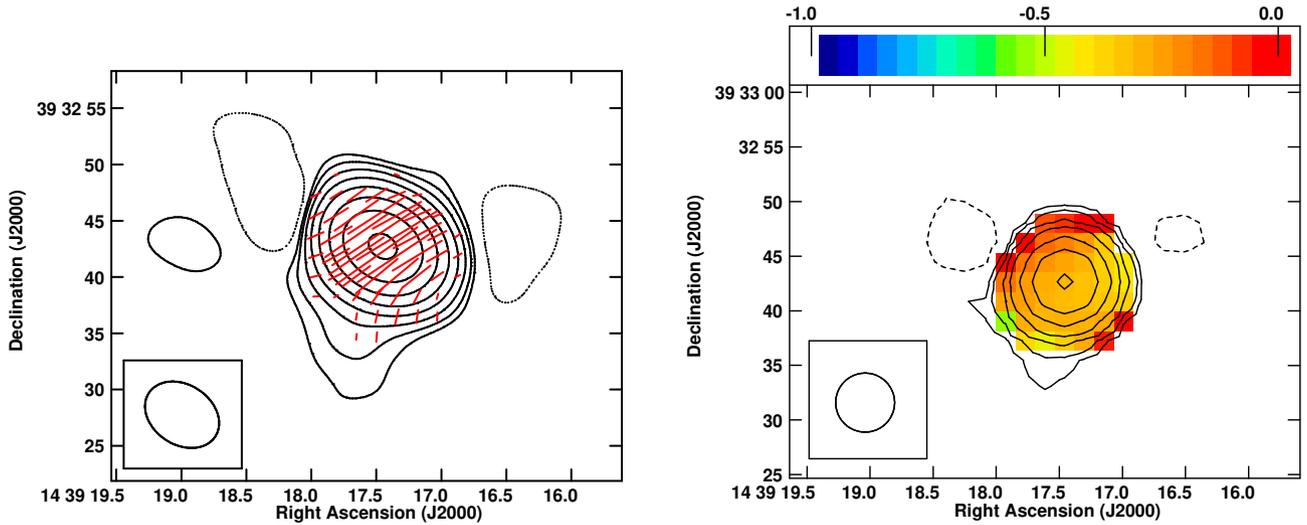

\centering
\begin{subfigure}{.5\linewidth}
  \centering
  \includegraphics[width=.6\linewidth, trim = 100 35 100 100]{PG1437_PPOL.eps}
  
\end{subfigure}%
\begin{subfigure}{.5\linewidth}
  \centering
  \includegraphics[width=.6\linewidth, trim = 100 17 100 0]{PG1437_spix2.eps}
  
\end{subfigure}
\caption{(Left) uGMRT contour image of the BL Lac object PG1437+398 at 650~MHz with EVPA vectors superimposed as red ticks. The length of the ticks are proportional to polarized intensity. The length of the ticks is proportional to polarized intensity: $10 \arcsec$ length corresponds to $1.67$~Jy~beam$^{-1}$. The contour levels are $8.57 \times 10^{-3} \times (\pm 0.5, 1, 2, 4, 8, 16, 32, 64)$~Jy~beam$^{-1}$. {The beam is $7.06 \times 5.32$ at a PA=$56.63\degr$.} (Right) 650~MHz uGMRT - 1.45~GHz VLA spectral index image in colour superimposed by uGMRT contours where the image is convolved with VLA beam. The contour levels are $8.67 \times 10^{-3} \times (0.35, 0.7, 1.4, 2.8, 5.6, 11.25, 22.5)$~Jy~beam$^{-1}$. {The beam is $5.40 \times 5.40$ at a PA=$0\degr$.}}
\label{fig7:Pg1437}
\end{figure*}

\subsubsection{PG1418+546 a.k.a. OQ +530} This source, also known as OQ530, is a radio-selected BL Lac and a member of the 1 Jy BL Lac sample \citep{Kuehr1981}. It is highly polarized at optical and near-infrared frequencies \citep{Mead1990}. During the periods when the source displays high polarization \citet{Marchenko1985} have found that this source might have a preferred polarization direction of near $\chi \sim$120$\degr$. This blazar is intraday optically variable \citep{Heidt1996} and has also displayed optical microvariability \citep{Miller1991}. It is an LSP blazar \citep{Lister2018, Ajello2022}.

The VLBI jet extends to the southeast in a structural PA $= 130\degr$ \citep{Gabuzda1996}. Both the VLBI core and two VLBI jet components, one compact and barely resolved from the core while the other more extended, display polarization. Polarization in the VLBI core is aligned with jet PA; the polarization in the inner knot is less well aligned with jet PA. There is no clear relation between the polarization in the outer region of the jet and the jet direction. Optical polarization position angle $\mathrm{EVPA_{opt}}$ is roughly aligned with VLBI jet polarization PA $\mathrm{EVPA_{jet}}$ in the inner knot, roughly perpendicular to $\mathrm{EVPA_{jet}}$ in the outer jet, and bears no obvious relation to the VLBI $\mathrm{EVPA_{core}}$ \citep{Gabuzda1996}. \citet{Lister2021} have provided a mean jet PA with 15~GHz VLBA observations as $+130\degr$. The stacked 15~GHz VLBA image shows EVPA parallel to the jet direction in the core and somewhat perpendicular but not well aligned in the other VLBI jet components \citep{Pushkarev2023}.

An asymmetric kpc-scale halo towards one side with a component about $32\arcsec$ to the west of the core is visible in the 1.6~GHz VLA image \citep{Murphy1993}. The diffuse halo was also seen in the 1.4~GHz VLA D-array image \citep{Cassaro1999}. This component appears to be a terminal hotspot making the BL~Lac morphology FRII-like. The kpc-scale lobe was found to be oppositely oriented to the single-sided jet visible in the VLBI images, which extends to the southeast to a depth of about 25 mas in the 5~GHz VLBA image \citep{Helmboldt2007}. The morphology of the 15~GHz VLBA image \citep{Lister2016} was consistent with that of the 5~GHz image. Superluminal motion or rapid structural change is not seen in the VLBA jet. Based on data collected at 15~GHz over an 18-year baseline from 1995 to 2013, a maximum jet speed of 0.93$\pm$0.27c has been obtained \citep{Lister2016}.

Our uGMRT 650~MHz image (Figure \ref{fig5:PG1418}) also shows a polarized compact core and a hotspot to the west, consistent with an FRII morphology. The spectral index image shows both the core and the hotspot to be relatively flat. The inferred B-field is indicative of shocked regions with the fields aligned with the edges in the hotspot region. The B-fields are perpendicular in the core with respect to the VLBI jet direction.

\subsubsection{PG1424+240 a.k.a. OQ +240} 
This source was identified as a blazar by \citet{Impey1988}. It was originally classified as a white dwarf in the PG sample but \citet{Fleming1993} used ROSAT All Sky Survey (RASS) to identify PG 1424+240 as a BL Lac object. It was first detected at VHE by VERITAS \citep{Acciari2010} and subsequently verified by MAGIC \citep{Aleksic2014}. At VHE $\gamma$-rays, its spectral energy distribution is significantly attenuated, which is consistent with distant sources. Several new redshift estimations have constrained it to the range 
0.6 $<$ z $<$ 1.3 \citep[e.g. see][]{Paiano17}. This places PKS 1424+240 in the very unusual position of probably being the most distant blazar that has been detected at VHE \citep{Rovero2016}. It is an HSP blazar \citep{Ajello2022}.

The VLBA images at 2 and 8~GHz show an unresolved source \citep{Fey1997}. Multi-epoch VLBI observations at 5~GHz, do not clearly resolve this source. However, the jet clearly extends to the southeast with a PA of about $\sim -150\degr$ \citep{Wu2007}. This jet is a little misaligned with the VLA image. The MERLIN archival data shows a very compact source, and the jet direction is not evident. The MOJAVE survey VLBA images of this source at 15, 23 and 43~GHz show a jet with a mean PA of $144\degr$ \citep{Lister2018,Lister2021}. \citet{Lister2021} have provided a mean jet PA with 15~GHz VLBA observations as $+144\degr$. The stacked 15~GHz VLBA image shows EVPA parallel to the jet direction in the core \citep{Pushkarev2023}. The VLA image of this BL~Lac shows a collinear jet/lobe structure with roughly co-linear jets extending to the north, PA of $\sim -10\degr$, and south PA of $\sim -175\degr$ \citep{Rector2003}. 

Our 650~MHz uGMRT image (Figure \ref{fig6:PG1424}) shows a morphology resembling a double-lobed Seyfert galaxy \citep[e.g., see][]{Gallimore2006,Das2021}. The core and lobes are both polarized. The spectral index image shows both the core to be relatively flat and then the spectra steepens towards the lobes. We find that the magnetic field is parallel to the lobe directions and also parallel to the VLBI jet direction in the core.

\subsubsection{PG1437+398} 
This object was identified as a BL~Lac object by \citet{Bauer2000,Piranomonte2007}. It is an HSP blazar \citep{Paiano2017A}. It belongs to the RGB sample of intermediate BL Lacs \citep{Laurent1999}. A possible parsec scale jet towards $\sim +45\degr$ is seen in the 8.3~GHz VLBI image in the Radio Fundamental Catalog\footnote{\url{http://astrogeo.org/rfc/}} {\citep[RFC; Petrov \& Kovalev, in preparation, as noted in][]{Petrov2019}}. The LOFAR image by \citet{Mooney2021} shows an unresolved structure with a possible extension towards the east. There is a double radio source to the northwest of this source at a distance of $\sim400$~kpc. There is a third component near this source at 14h37m19.0s, +39{\degr}45{\arcmin}35{\arcsec} (epoch = B1950; 1460~MHz flux is 35 mJy) coinciding with a 15 mag compact object, perhaps a galaxy \citep{Vigotti1989}.

Our 650~MHz uGMRT image (Figure \ref{fig7:Pg1437}) shows a core-halo morphology with indications of a kpc scale jet-like extension in the southern direction. The core is polarized and the inferred B-fields there are perpendicular to the kpc jet axis. The spectral index image shows the core to be flat.

\begin{figure*}
\centering
\begin{subfigure}{.5\linewidth}
  \centering
  \includegraphics[width=.6\linewidth, trim= 150 240 150 240]{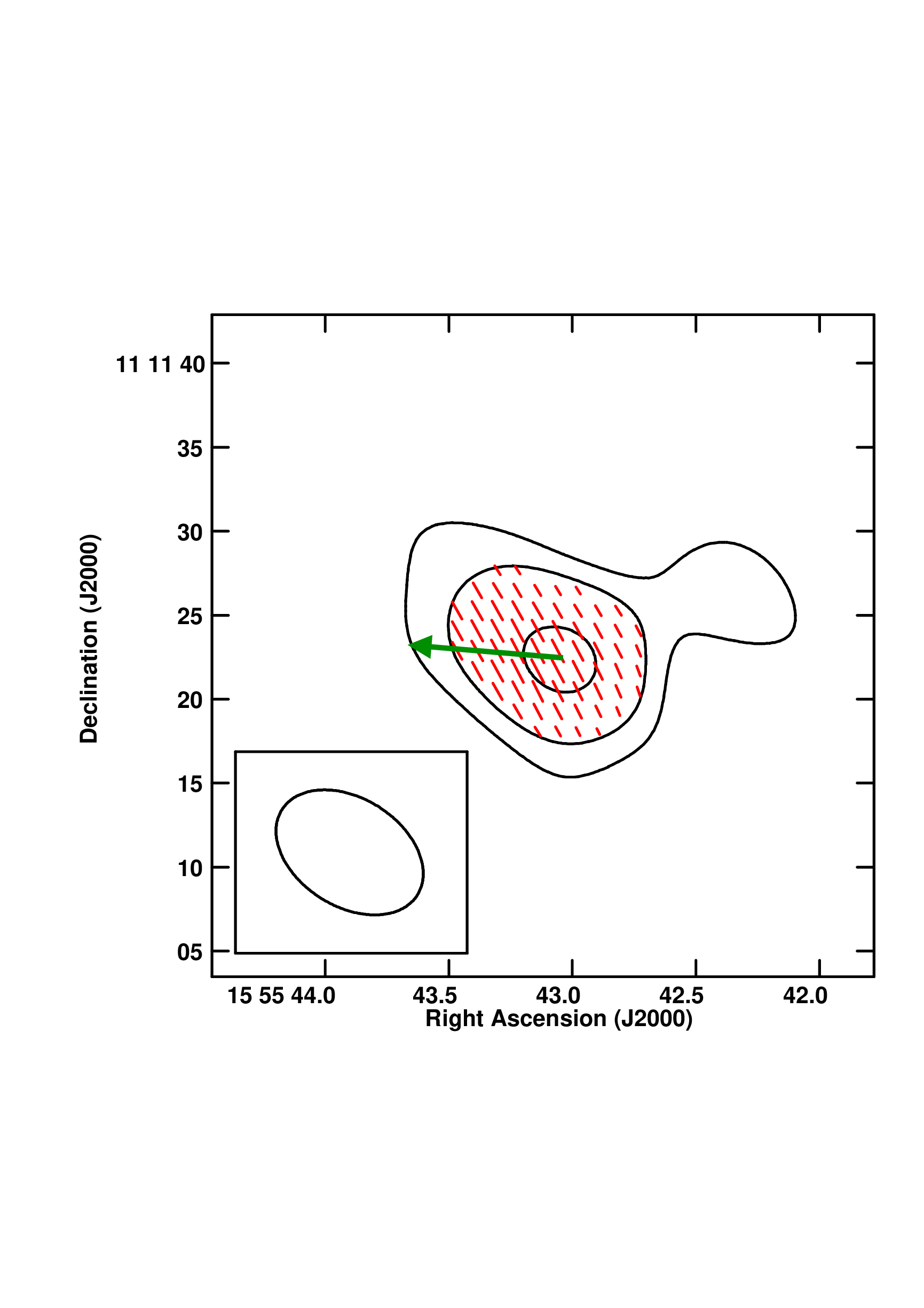}
  
\end{subfigure}%
\begin{subfigure}{.5\linewidth}
  \centering
  \includegraphics[width=.6\linewidth, trim= 90 50 90 210]{PG1553_alpha_clr.eps}
  
\end{subfigure}
\caption{(Left) uGMRT contour image of the BL Lac object PG1553+113 at 650~MHz with global RM corrected EVPA vectors superimposed as red ticks and VLBI jet direction shown by the green arrow. The length of the ticks is proportional to polarized intensity: $5 \arcsec$ length corresponds to $4.167$~mJy~beam$^{-1}$. The contour levels are $1.61 \times 10^{-3} \times (22.5, 45, 90)$~Jy~beam$^{-1}$. The beam is $9.53 \times 6.43$ at a PA=$57.82\degr$. (Right) 650~MHz uGMRT - 1.45~GHz VLA spectral index image in colour superimposed by contours of uGMRT image with the VLA image convolved with the uGMRT beam. The contour levels are $1.61 \times 10^{-3} \times (\pm 22.5, 45, 90)$~Jy~beam$^{-1}$. The beam is $9.53 \times 6.43$ at a PA=$57.82\degr$.} 
\label{fig8:PG1553}
\end{figure*}

\begin{figure*}
\centering
\begin{subfigure}{.5\linewidth}
  \centering
  \includegraphics[width=.6\linewidth, trim = 310 310 310 420]{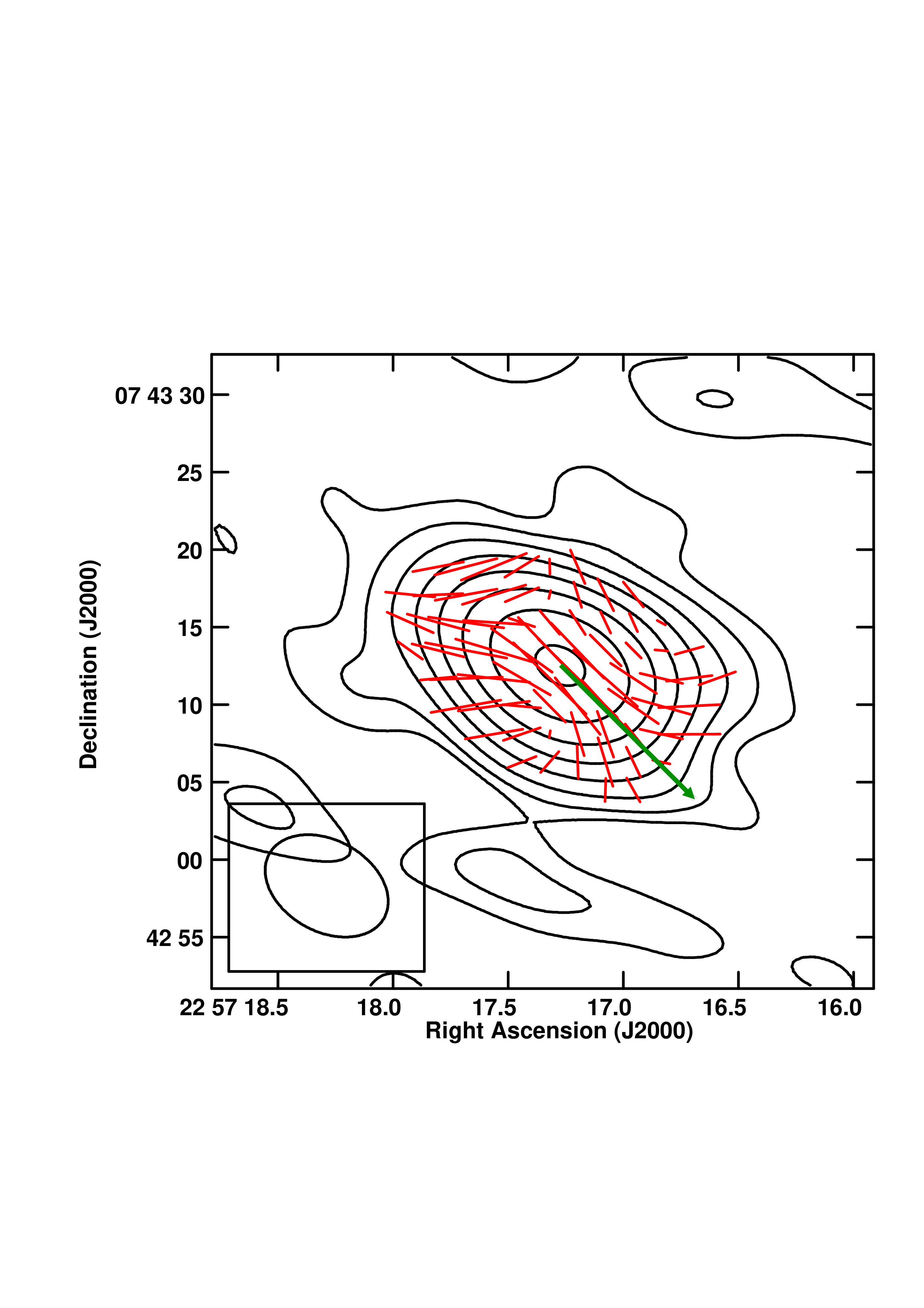}
  
\end{subfigure}%
\begin{subfigure}{.5\linewidth}
  \centering
  \includegraphics[width=.6\linewidth, trim = 100 17 100 0]{PG2254_new_alpha_clr.eps}
  
\end{subfigure}
\caption{(Left) uGMRT contour image of the BL Lac object PG2254+075 at 650~MHz with global RM corrected EVPA vectors superimposed as red ticks and VLBI jet direction shown by the green arrow. The length of the ticks are proportional to polarized intensity: $5 \arcsec$ length corresponds to $10.4$~mJy~beam$^{-1}$. The contour levels are $2.92 \times (1.4, 2.8, 5.6, 11.25, 22.5, 45, 90)$~mJy~beam$^{-1}$. {The beam is $8.48 \times 5.89$ at a PA$=60.66\degr$.}
(Right) 650~MHz uGMRT - 1.45~GHz VLA spectral index image in colour superimposed by contours of uGMRT image convolved with the VLA beam. The contour levels are $3.11 \times (0.7, 1.4, 2.8, 5.6, 11.25, 22.5, 45, 90)$~mJy~beam$^{-1}$. {The beam is $16.82 \times 12.52$ at a PA$=3.25\degr$.}}
\label{fig9:PG2254}
\end{figure*}

\subsubsection{PG1553+113} This is an HSP Fermi-detected blazar \citep{Lister2018,Ajello2022}. It shows quasi-periodic high-energy variability and has been detected at MeV/GeV energies by the Fermi-LAT \citep{Abdo2009,Ackermann2015}. The RoboPol blazar monitoring program and the NOT have found an average polarization of about $3.4\%$ and an average optical EVPA of $87\degr$ \citep{Hovatta2016}.

The VLBA 5~GHz image shows a jet extending to the northeast \citep[PA of $\sim +48\degr$;][]{Rector2003}. Using VLBA at 15, 24, and 43 GHz \citep{Lico2020} found a core-dominated structure with a limb-brightened $\sim 1.5$ mas jet towards the northeast. \citet{Lister2021} have provided a mean jet PA with 15~GHz VLBA observations as $+85\degr$. The stacked 15~GHz VLBA image shows EVPA not well aligned/oblique to the jet direction in the core and EVPA parallel to the jet direction further away from core \citep{Pushkarev2023}. Beyond 20 parsecs the jet is very faint making it difficult to determine its opening angle or whether it is bent. No continuous jet from the core is detected in a deep VLA image; however, a faint lobe is detected south of the core, with a weak hot spot at a PA of $\sim +160\degr$, giving a large misalignment angle of $\sim 112\degr$ if we assume the VLBA jet is related to the southern lobe \citep{Rector2003}. 

Our 650~MHz uGMRT image (Figure \ref{fig8:PG1553}) shows a compact core and a possible halo surrounding it. The core is polarized. The B-fields are not neatly aligned with the VLBI jet direction in the core, it is tending towards more parallel than perpendicular. The spectral index image shows both the core to be relatively flat with the spectra steepening towards the outer halo. We note that the uGMRT observations of PG1553+113 have a relatively poor signal-to noise ratio compared to the other sources and would benefit from more sensitive data.

\subsubsection{PG2254+075 a.k.a. OY +091} This is a radio-selected BL Lac belonging to the 1 Jy sample, as well as the MOJAVE sample. It is an LSP blazar \citep{Giroletti2004,Ajello2022}. The Fermi $\gamma$-ray point source associated with this source is offset from its radio core by $6.235\arcmin$ \citep{Ajello2022}. The VLBA 2 and 8~GHz total flux density is 350 mJy \citep[][]{Fey2000} and the jet is oriented towards the west. \citet{Lister2021} have provided a mean jet PA with 15~GHz VLBA observations as $-135\degr$. The stacked 15~GHz VLBA image shows EVPA parallel to the jet direction in both the core and jet \citep{Pushkarev2023}. 

This source has one of the brightest cores with VLA {C-array,~5~GHz} luminosity (log $L=26.1$~W~Hz$^{-1}$) \citep{Giroletti2004}. Little extended emission is present on the arcsecond scale, both in the A and C configuration images of the VLA \citep{Giroletti2004}. A comparison to the parsec-scale total flux density indicates that some emission originates on intermediate scales.

Our 650~MHz uGMRT (Figure \ref{fig9:PG2254}) image shows a core-halo morphology. The core polarization shows B-fields perpendicular to the VLBI jet. The core and lobe show clearly differing polarization structures with the inferred B-fields in the lobes appearing to be largely tangential. This core-lobe polarization structure is reminiscent of a `spine-sheath structure' with the `sheath' displaying primarily toroidal magnetic fields \citep[see IIIZw2 in][]{Silpa2021}. The spectral index image shows the core to be relatively flat with the spectra steepening in the halo/lobes. 

\begin{table*}
\centering
\caption{\label{tab5}The PG BL Lac Sample Properties}
\begin{tabular}{c|c|c|c|c|c|c|c|c|c|}
\hline
S.No. & Name & 
 $\mathrm{\log_{10} (M_{BH} / M_{\sun})}$ & $\delta$ & Ref & $\bar{Q}$ ($\mathrm{\times10^{42}~erg~s^{-1}}$) & $L_\gamma$ ($\mathrm{erg~s^{-1}}$) & Photon index &Ref & $\mathrm{\log_{10} (\nu_s / Hz)}$
\\ \hline
1 & {PG0851+203} &  8.5 &17.87 & 1 & 210 & ($1.62\pm0.04$)$\times10^{46}$ & $2.21\pm0.01$ &2 & 13.24\\
2 & {PG1101+384} &  8.23 &1.99 & 1 & 7 & ($8.4\pm0.1$)$\times10^{44}$ & $1.78\pm0.005$ &2 & 16.22\\
3 & {PG1218+304} &  8.47 &2.60 & 1 & 9 & ($4.0\pm0.1$)$\times10^{45}$ & $1.72\pm0.02$ &2 & 16.27\\
4 & {PG1418+546} &  8.74 &11.38 & 1 & 62 & ($7.3\pm0.4$)$\times10^{44}$ & $2.33\pm0.03$ &2 & 13.68 \\
5 & {PG1424+240} &  6.42 &2.66 & 1 & 988 & ($1.47\pm0.03$)$\times10^{47}$ & $1.82\pm0.01$ &2 & 15.29 \\
6 & {PG1437+398} &  8.95 &1.87 & 1 & 165 & ($2.2\pm0.2$)$\times10^{45}$ & $1.96\pm0.05$ &2 & 15.86\\
7 & {PG1553+113} &  7.25 &5.07 & 1 & 120 & ($6.6\pm0.1$)$\times10^{46}$ & $1.68\pm0.01$&2 & 15.59\\
8 & {PG 2254+075} &  8.85 & 8.85 & 1 & 43 & ($3.4\pm0.5$)$\times10^{44}$ & $2.21\pm0.09$&2 & 12.78\\
\hline
\multicolumn{10}{l}{Note. Column (1): Serial Number. Column (2): PG names. Column (3): Black hole masses. Column (4): Doppler factor. }\\
\multicolumn{10}{l}{Column (5): References for black hole masses and Doppler factor. Column (6):  Bulk radio jet kinetic power. Column (7):}\\
\multicolumn{10}{l} {High Energy luminosity (0.1 - 100 GeV). Column (8): Photon index for the PowerLaw fit at high energies (0.1$-$100 GeV).}\\
\multicolumn{10}{l}{Column (9): References for High Energy Luminosity and Photon index. Column (10): Synchrotron peak frequency}\\
\multicolumn{10}{l}{References- 1: \citet{Wu_2009} 2: \citet{Abdollahi2022}}\\
\end{tabular}
\end{table*}

\subsection{Jet Properties through Global Correlations}
\begin{figure*}  
\centerline{
\includegraphics[width=9.5cm]{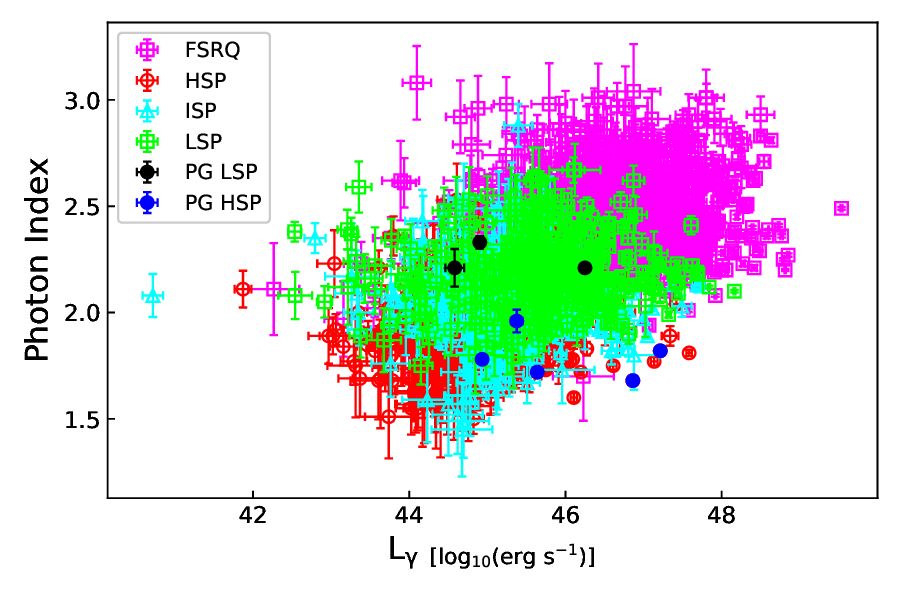}}
\caption{Distribution of $\gamma$-ray luminosity and photon indices of the BL Lacs and FSRQs reported in the 4FGL-DR3 catalog \citep{Ghisellini2010,Bhattacharya2016} with the PG BL Lac objects overlaid in black and blue symbols.}

\label{corr4}
\end{figure*}

\begin{figure*}  
\centerline{
\includegraphics[width=9.8cm,trim=-20 0 0 0]{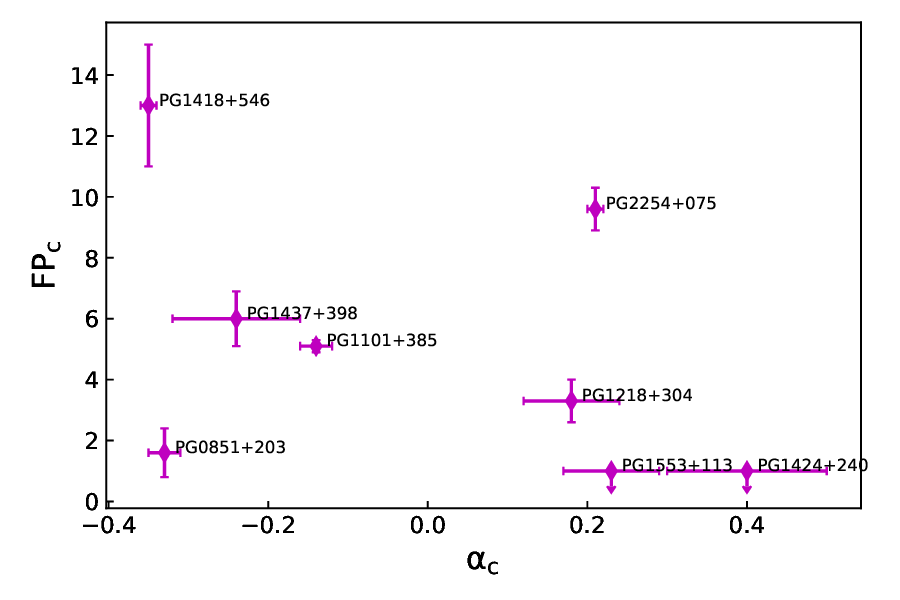}
\includegraphics[width=9.8cm]{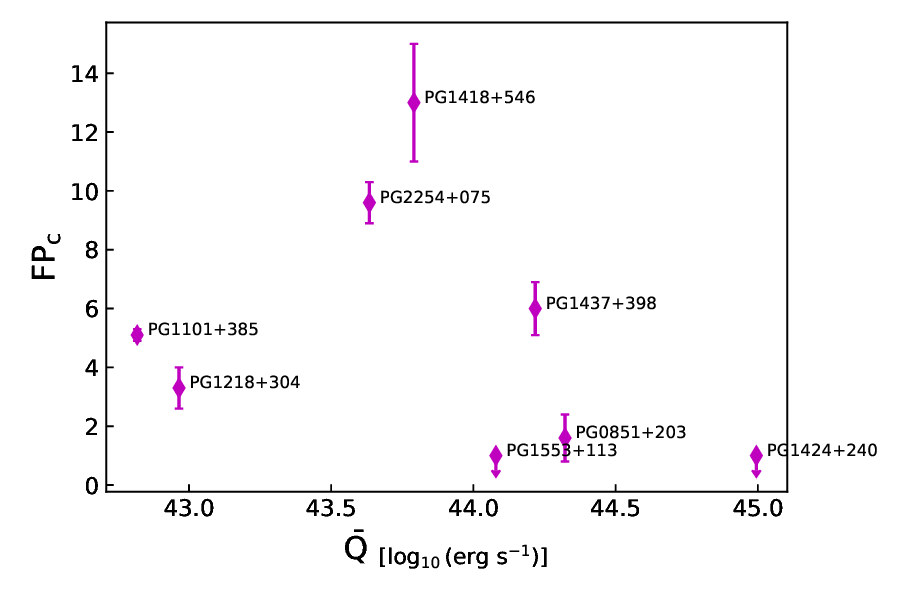}
}
\caption{(Left) Core fractional polarization versus core spectral index for the BL Lac objects. (Right) Core fractional polarization versus jet power
}
\label{corr1}
\end{figure*}

\begin{figure*}  
\centerline{
\includegraphics[width=9cm]{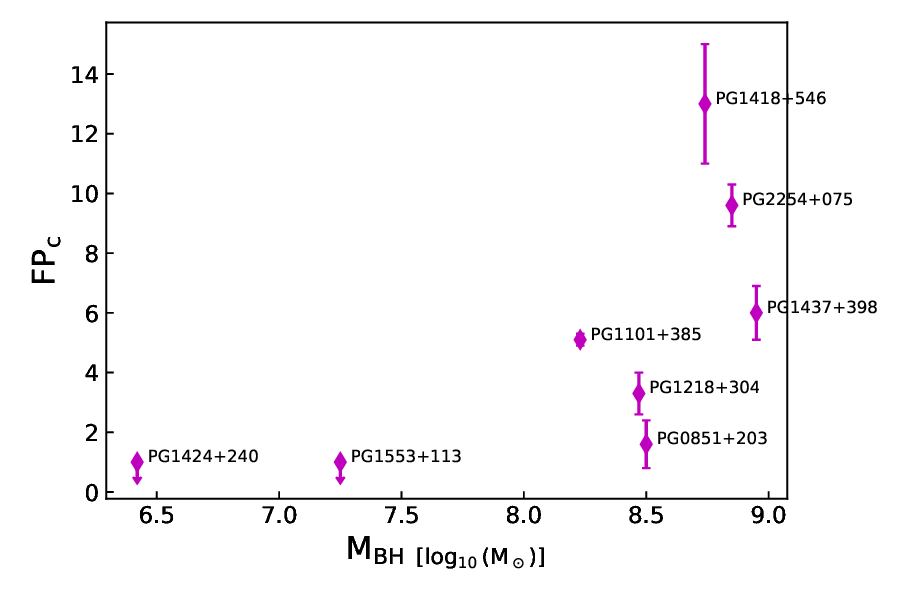}
\includegraphics[width=9cm]{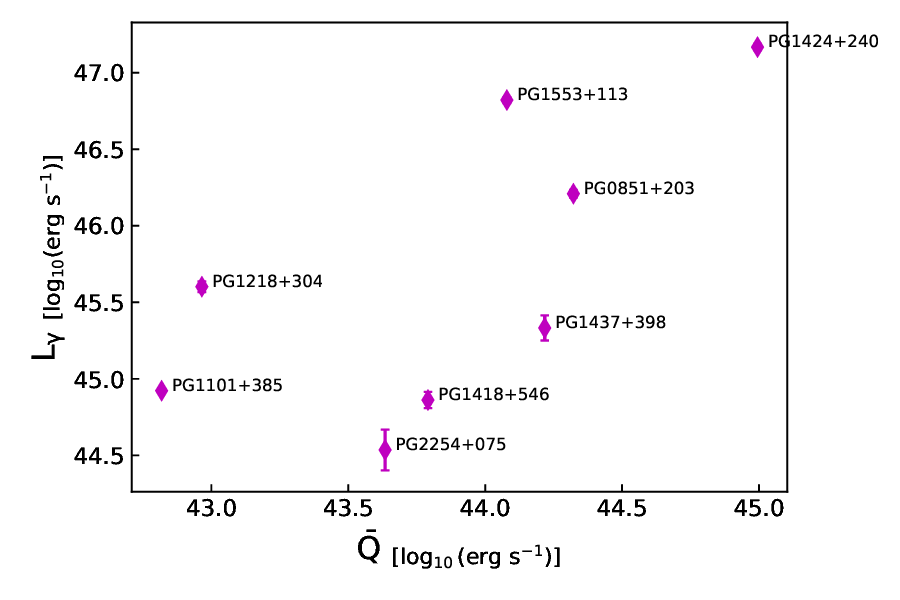}}
\caption{(Left) Core fractional polarization versus black hole mass (in units of $\log_{10} M_\odot$) for the BL Lac objects. (Right) Log of 1-100 GeV $\gamma$-ray luminosity with the log of jet power in erg~s$^{-1}$.}
\label{corr2}
\end{figure*}

The high energy (HE) characteristics of blazars show differing behavior in the case of BL Lac objects and quasars \citep{Ghisellini2017}. The broad line region clouds and the dusty torus are thought to be sources of seed photons that lead to a high rate of radiative cooling in the case of quasars \citep{Ghisellini2013, Ghisellini2017}. The origin of the HE emission in blazars in terms of its nature and location is still not well understood \citep{Madejski2016, Hovatta2019}. Since all the PG BL~Lac objects are strong $\gamma$-ray emitters, we have gathered the HE data from the literature as well as estimated jet powers from the low-frequency radio flux density for all the sources. In Figure~\ref{corr4} we have plotted the HE emission of the PG BL~Lacs and put them in context with a larger blazar sample; the distribution of $\gamma$-ray luminosities and photon indices of Fermi blazars from the most recent catalog \citep[4FGL-DR3 and 4LAC-DR3; ][]{Abdollahi2022, Ajello2022}, with the sources in our sample marked is shown. It is clear that the PG BL Lacs are typical of the general population of BL~Lac objects.

In order to better understand the HE emission, we look for relations with kpc-scale polarization properties which are in turn related to the order of the large-scale B-fields as well as the depolarizing environment. We have estimated the long-term time-averaged bulk radio jet kinetic power $\bar{Q}$ using the radio luminosity at 151~MHz as a surrogate for the luminosity of the radio lobes by using the relation given by \citet{Punsly2018}, 
$$\bar{Q} = 3.8 \times 10^{45} \it{f}~L_{151}^{6/7}~\mathrm{erg~s^{-1}}$$ where $L_{151}$ is radio luminosity at 151~MHz in units of $10^{28}$~W~Hz$^{-1}$~sr$^{-1}$ and $f\approx15$ \citep{Blundell2000}. We obtained the $L_{151}$ from the TGSS survey flux densities \citep{Intema2017} using the relation, $L_{151} = [ D_{L}^2~F_{151}]/[(1+z)^{(1+\alpha)}]$~W~Hz$^{-1}$~sr$^{-1}$, \footnote{We note that there is no $4\pi$ factor because of the anisotropic jet emission, e.g., see \citet{Peacock1999}.} where we have used the typical average $\alpha = -0.7$ for the extended emission.

We report $\bar{Q}$ without errors as systematic uncertainties in the estimation methods dominate the statistical uncertainty in the data. Black hole masses ($M_{BH}$) and Doppler factor ($\delta$) do not have reported errors in \citet{Wu_2009} due to the same reason. High energy luminosity ($0.1-100$~GeV) ($L_\gamma$), photon index, and synchrotron peak frequency $\nu_s$ have been reported from FERMILPSC - Fermi LAT 12-Year Point Source Catalog \citep[4FGL-DR3;][]{Abdollahi2022} and the Fourth LAT AGN Catalog \citep[4LAC-DR3;][]{Ajello2022}. The errors in these values are plotted as reported in the catalog or calculated from reported errors in the flux values. 

\begin{figure*}  
\centerline{
\includegraphics[width=9cm]{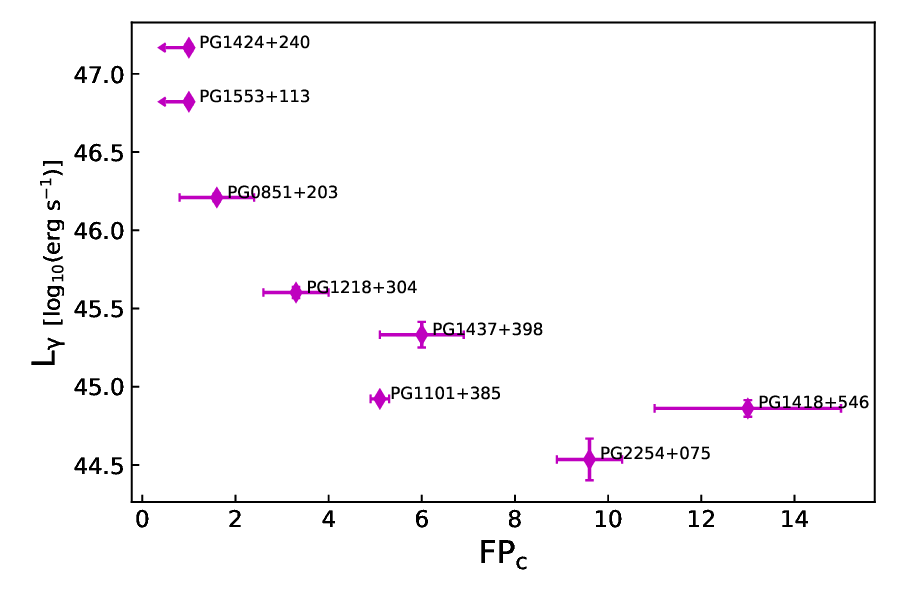}
\includegraphics[width=9cm]{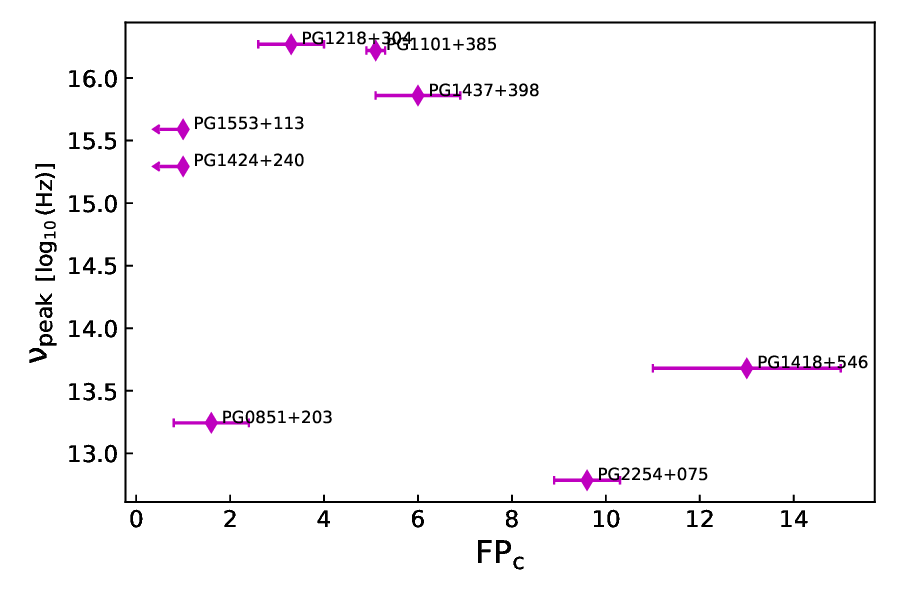}}
\caption{(Left) Log of high energy $\gamma$-ray luminosity versus core fractional polarization. (Right) Log of synchrotron peak frequency vs core fractional polarization}
\label{corr3}
\end{figure*}

We find a marginal positive correlation between the black hole masses and the uGMRT core fractional polarization ($FP_c$) (p-value = 0.061). This marginal correlation between core fractional polarization and black hole mass was also found in our sample of PG quasars \citep{Baghel2023}. This however has some caveats as we discuss in Section \ref{Disc}. We also find a statistically significant negative correlation between $FP_c$ and the $L_\gamma$ (p-value = 0.0041). We found no significant correlation between core spectral index $\alpha$ and $FP_c$ (p-value = 0.105), $\bar{Q}$ and $FP_c$ (p-value = 0.261), $L_\gamma$ and $\bar{Q}$ (p-value = 0.108), and $\nu_s$ and $FP_c$ (p-value = 0.533). However, the $\nu_s$ and $FP_c$ plot shows that all the HSP PG BL Lacs have lower $FP_c$ whereas LSPs are more spread out. 

The radio core prominence, $R_c$, is a statistical indicator of beaming and thereby orientation \citep[e.g.,][]{Orr1982,Kharb2004}. It is defined as the ratio of the core flux density (expected to be dominated by the relativistic jet and therefore beamed) to the extended flux density (expected to be largely unbeamed): $\mathrm{R_c = \frac{Core~flux~density}{(Total~flux~density - Core~flux~density)}{(1+z)^{1+\alpha}}}$, where $\alpha=-0.7$. We found no significant correlation between R$_c$ and $FP_c$, M$_{\text{BH}}$, photon index, or $L_\gamma$, suggesting that these quantities are not related to orientation. Interestingly, we find no correlation between the R$_c$ and the Doppler factor $\delta$. This could be explained if the  orientation of the jet changes appreciably between parsec and kpc scales \citep[e.g.,][]{Kharb10}.
 
\section{Discussion} \label{Disc}
Our uGMRT observations have detected linear polarization at 650~MHz in all the 8 BL~Lac objects belonging to the PG `blazar' sample. Polarization is detected in the inner core-jet regions for sources exhibiting a core-halo radio morphology on kpc-scales. Two BL~Lac objects, viz., PG0851+203 and PG1418+546, show the presence of terminal hotspots on both or one side of the radio core, making them similar to pole-on FRII radio galaxies. When hotspots are observed, polarization is detected in them with the inferred B-fields resembling terminal-shock-like features with compressed B-fields perpendicular to the jet directions. 

These results are broadly similar to what we have observed in the PG quasars which additionally show a large fraction of hybrid morphology or restarted sources \citep{Baghel2023}. While keeping in mind the small number statistics, it appears that the PG BL~Lacs show less diversity in their radio properties compared to the PG quasars. It is interesting that based on a large sample of FR radio galaxies studied in the LOFAR Two-Metre Sky Survey (LoTSS), \citet{Mingo2019} found that FRII sources spanned both above and below the FR divide in terms of their radio luminosity with a large fraction of FRIIs being of lower luminosity. 

We find no correlation between the mean VLBI jet PA or the 15~GHz VLBA stacked EVPA direction in the radio core with the 650~MHz uGMRT EVPA direction in the core (see Table \ref{tab4}). This is likely to be due to largely different spatial scales sampled with the uGMRT data and the VLBA. In their observations of the kpc-scale radio morphology of BL~Lac objects with LOFAR, \citet{Mooney2021} found extended emission in $\sim67\%$ (66/99) of the BL~Lacs. A minority of the objects were also found to be core-dominated sources with smaller spatial extents like FR0 sources. FR0s are a relatively new class of faint yet abundant AGN with no extended kpc-scale radio emission, distinguishing them from FRIs and FRIIs \citep{Ghisellini2011b,Baldi2016}. Their core luminosity is similar to that of FRI radio sources but the extended emission is weaker by a factor of $\sim$100 \citep{Baldi2009,Cheng2018}. \citet{Massaro2020} have suggested that the environments of BL~Lac objects are more like FR0s than FRIs.  We find two BL~Lac objects, viz., PG1218+304 and perhaps PG1553+113, which do not show clear signs of extended emission and could in principle be similar to FR0s in morphology. 

\citet{Smierciak2022} have suggested that the extended features observed in some of the BL~Lac objects like Mrk421 could be due to jet re-orientation out of the sky plane. The unbeamed large radio structures lie in the sky plane while the kpc-scale jet is moving towards us and is responsible for the blazar nature of these sources. This explanation is likely to hold for several BL~Lacs in our sample like PG1418+546 and PG1424+240 where the VLBI jet direction is not closely aligned with the kpc-scale jets or lobes. This is also consistent with us not finding any significant correlations between the radio core prominence parameter R$_c$ and other parameters. For instance, the lack of a correlation between the R$_c$ and the Doppler factor $\delta$ is consistent with the suggestion of the orientation of the jet changing between parsec and kpc scales. BL~Lac jets likely undergo multiple re-orientations as they propagate outwards, possibly due to interaction with the surrounding medium. However, the presence of polarized emission on arcsec-scales indicates that the amount of mixing of thermal plasma with synchrotron plasma is inadequate to completely depolarize the emission by the internal depolarization mechanism.

In an effort to place the PG BL~Lac objects in the context of larger samples of BL~Lacs, we find that the two sources in the sample showing FRII-like morphologies (i.e., hotspots) are LSP sources with steeper $\gamma$-ray photon indices, viz., PG0851+203 and PG1418+546. From the $\gamma$-ray luminosity distribution plot of HSP, ISP, LSP and FSRQ in the second Fermi catalog \citep{Ghisellini2010,Bhattacharya2016}, it is found that the $\gamma$-ray behavior of LSPs (luminosities and photon indices) is closer to those of FSRQs. These two sources also have higher Doppler beaming factors (Table \ref{tab5}) than the rest. Typically, FSRQs and LSP BL Lacs have a higher Doppler beaming factor than HSP BL Lacs \citep{Lister2011}. These results are consistent with their observed hybrid morphologies as well. This provides support to the current `blazar envelope' picture \citep[e.g.,][]{Meyer2011} where the LSP BL~Lacs have properties intermediate between the quasars and HSP BL~Lacs. 
These results are consistent with their observed hybrid morphologies as well, and in slight friction with blazar sequence scenario \citep{Fossati1998, Ghisellini2017} where BL~Lac objects and FSRQs are intrinsically different.
 
The nature of this transition and even the existence of both the blazar sequence and the blazar envelope is, however, still very much debated \citep{Meyer2012,Keenan2021} and does not explain all the observed sources \citep{Cerruti2017}.

The correlations found between core fractional polarization $FP_c$ and black hole masses in both the PG quasars and PG BL~Lacs suggest that a higher black hole mass is associated with greater ordered magnetic field that is sustained till kpc-scales. However, we note that the black hole mass estimates of the BL~Lacs are not very reliable as they have been derived using single-band optical luminosities and an empirical correlation, which is inconsistent with the recent correlations of e.g., \citet{Kormendy2013} that have a different zero point and use model radii from reverberation mapping, among other differences.
However, we continue to use these values as they are the only uniformly estimated ones available for our entire sample of BL~Lac objects. We note however, that if we exclude the unusually low black hole masses of PG1553+113 and PG1424+240, we find no correlation between $FP_c$ and black hole masses (p-value = 0.719).

We find an anti-correlation between $FP_c$ and the $L_\gamma$ and an absence of a correlation in $FP_c$ and $\nu_s$. However, our HSP BL~Lacs cluster towards the lower end of $FP_c$ while LSP BL~Lacs show a wider range in $FP_c$. \citet{Lister2011} have found a similar trend in SED peaks by comparing the parsec scale radio fractional core polarization in MOJAVE blazars. Therefore, even though the number of PG BL~Lacs is small, their behavior is similar to the general trend observed for a larger sample of blazars. This is a possible reason for the anti-correlation observed in the $L_\gamma$ versus $FP_c$ plot (Fig. \ref{corr3}, left panel). Moreover, the $\gamma$-ray luminosity sampled by Fermi does not give us a full picture of the high energy emission in these blazars as they sample different portions of the second SED peak due to the fixed observing frequency range. Sources that have their inverse Compton peak within this range tend to show higher luminosities \citep{Lister2015}, producing the observed trend.

\section{Conclusions}
We have presented here uGMRT band 4 ($\sim650$~MHz) polarization images of 8 BL~Lac objects belonging to the Palomar-Green (PG) `blazar' sample. These are the first-ever observations of these BL~Lac objects at low radio frequencies ($\le1$~GHz) in polarized light at high sensitivity. We summarize the main results below: 

\begin{enumerate}
\item Nearly two-thirds of the BL~Lacs reveal core-halo radio structures (PG1424+240 shows a FR type I-lobe-like structure) with most of the polarization detected in the inner core-jet regions. PG1101+385 and PG2254+075 show a `spine-sheath structure' in polarization. The core-halo and `spine-sheath' structures are consistent with the Unified Scheme suggestion that BL~Lacs are the pole-on beamed counterparts of FRI radio galaxies. Fractional polarization ranges from $1-13$\% in the cores and $2-26$\% in the inner jets/lobes of the BL~Lacs. 

\item PG1418+546 and PG0851+203 (a.k.a. OJ287) show the presence of terminal hotspots similar to FRII radio galaxies. As these sources have been identified as low-spectrally peaked BL Lacs, their radio morphology being FRII or quasar-like supports the idea of a `blazar envelope' reported in the literature.

\item Compared to the PG quasars from the PG `blazar' sample \citep{Baghel2023} which showed several distorted, hybrid and restarted sources, BL~Lacs appear to be less diverse in terms of their radio properties. 

\item A comparison of the inferred core magnetic field structures on arcsec- or kpc-scales w.r.t. the VLBI jet direction does not reveal any preferred orientation, suggesting that if large-scale ordered B-fields do exist, they do so on scales smaller than probed by the current observations. 

\item The presence of polarized emission on arcsec-scales suggests that any mixing of thermal plasma with the synchrotron emitting plasma is not sufficient to fully depolarize the emission via the internal depolarization mechanism. 

\item  We found no significant correlation between the statistical indicator of orientation, R$_c$, and core fractional polarization, black hole mass, or high energy luminosity. We also found no correlation between the parsec-scale and kpc-scale jets of PG BL~Lacs. These results support the suggestion of re-orientation of jets as they propagate to kpc scales.

\item The marginal positive correlation found between core fractional polarization and black hole mass in both the PG quasars and PG BL~Lac objects suggests that a higher black hole mass is associated with more organised magnetic fields that are sustained till kpc-scales. 

\item We observe an anti-correlation between the core fractional polarization and $\gamma$-ray luminosity. However, this is due to the HSP BL~Lacs being clustered towards the lower linear core polarization levels, as opposed to the LSP BL~Lacs that show a wide range of core fractional polarizations \citep[see also][]{Lister2011}. 

Detailed broadband spectral and temporal studies of these sources, together with deep radio observations (with uGMRT, LOFAR), will provide a better understanding of the observed correlations and their underlying physical processes.

\end{enumerate}

\section*{Acknowledgements}
We thank the referee for their constructive suggestions that have improved the manuscript.
JB, PK, SG and SS acknowledge the support of the Department of Atomic Energy, Government of India, under the project 12-R\&D-TFR-5.02-0700. PK acknowledges the support of the Centre for Astrophysics | Harvard \& Smithsonian, as a visiting scientist. TH was supported by the Academy of Finland projects 317383, 320085, 322535, and 345899. E. L. was supported by the Academy of Finland projects 317636, 320045 and 346071. We thank the staff of the GMRT that made these observations possible. GMRT is run by the National Centre
for Radio Astrophysics of the Tata Institute of Fundamental Research. We acknowledge the support of the Department of Atomic Energy, Government of India, under the project 12-R\&D-TFR-5.02-0700. The National Radio Astronomy Observatory is a facility of the National Science Foundation operated under cooperative agreement by Associated Universities, Inc. This research has made use of the NASA/IPAC Extragalactic Database (NED), which is operated by the Jet Propulsion Laboratory, California Institute of Technology, under contract with the National Aeronautics and Space Administration. This research has made use of the VizieR catalogue access tool, CDS, Strasbourg, France (DOI : 10.26093/cds/vizier). The original description of the VizieR service was published in 2000, A\&AS 143, 23.

\section*{Data Availability}

The data underlying this article will be shared on reasonable request to the corresponding author. The GMRT data underlying this article can be obtained from the GMRT Archive (http://naps.ncra.tifr.res.in/goa/data/search) using the proposal ID: 42\_091, DDTC217 and DDTC273. 

\appendix

\bsp	
\label{lastpage}

\bibliographystyle{mnras}
\bibliography{main} 

\begin{thebibliography}{}
\makeatletter
\relax
\def\mn@urlcharsother{\let\do\@makeother \do\$\do\&\do\#\do\^\do\_\do\%\do\~}
\def\mn@doi{\begingroup\mn@urlcharsother \@ifnextchar [ {\mn@doi@} {\mn@doi@[]}}
\def\mn@doi@[#1]#2{\def\@tempa{#1}\ifx\@tempa\@empty \href {http://dx.doi.org/#2} {doi:#2}\else \href {http://dx.doi.org/#2} {#1}\fi \endgroup}
\def\mn@eprint#1#2{\mn@eprint@#1:#2::\@nil}
\def\mn@eprint@arXiv#1{\href {http://arxiv.org/abs/#1} {{\tt arXiv:#1}}}
\def\mn@eprint@dblp#1{\href {http://dblp.uni-trier.de/rec/bibtex/#1.xml} {dblp:#1}}
\def\mn@eprint@#1:#2:#3:#4\@nil{\def\@tempa {#1}\def\@tempb {#2}\def\@tempc {#3}\ifx \@tempc \@empty \let \@tempc \@tempb \let \@tempb \@tempa \fi \ifx \@tempb \@empty \def\@tempb {arXiv}\fi \@ifundefined {mn@eprint@\@tempb}{\@tempb:\@tempc}{\expandafter \expandafter \csname mn@eprint@\@tempb\endcsname \expandafter{\@tempc}}}

\bibitem[\protect\citeauthoryear{{Abdo} et~al.,}{{Abdo} et~al.}{2009}]{Abdo2009}
{Abdo} A.~A.,  et~al., 2009, \mn@doi [\apjs] {10.1088/0067-0049/183/1/46}, \href {https://ui.adsabs.harvard.edu/abs/2009ApJS..183...46A} {183, 46}

\bibitem[\protect\citeauthoryear{{Abdo} et~al.,}{{Abdo} et~al.}{2010}]{Abdo2010}
{Abdo} A.~A.,  et~al., 2010, \mn@doi [\apj] {10.1088/0004-637X/716/1/30}, \href {https://ui.adsabs.harvard.edu/abs/2010ApJ...716...30A} {716, 30}

\bibitem[\protect\citeauthoryear{{Abdollahi} et~al.,}{{Abdollahi} et~al.}{2022}]{Abdollahi2022}
{Abdollahi} S.,  et~al., 2022, \mn@doi [\apjs] {10.3847/1538-4365/ac6751}, \href {https://ui.adsabs.harvard.edu/abs/2022ApJS..260...53A} {260, 53}

\bibitem[\protect\citeauthoryear{{Acciari} et~al.,}{{Acciari} et~al.}{2010}]{Acciari2010}
{Acciari} V.~A.,  et~al., 2010, \mn@doi [\apjl] {10.1088/2041-8205/708/2/L100}, \href {https://ui.adsabs.harvard.edu/abs/2010ApJ...708L.100A} {708, L100}

\bibitem[\protect\citeauthoryear{{Ackermann} et~al.,}{{Ackermann} et~al.}{2015}]{Ackermann2015}
{Ackermann} M.,  et~al., 2015, \mn@doi [\apjl] {10.1088/2041-8205/813/2/L41}, \href {https://ui.adsabs.harvard.edu/abs/2015ApJ...813L..41A} {813, L41}

\bibitem[\protect\citeauthoryear{{Agudo}, {Marscher}, {Jorstad}, {G{\'o}mez}, {Perucho}, {Piner}, {Rioja}  \& {Dodson}}{{Agudo} et~al.}{2012}]{Agudo2012}
{Agudo} I.,  {Marscher} A.~P.,  {Jorstad} S.~G.,  {G{\'o}mez} J.~L.,  {Perucho} M.,  {Piner} B.~G.,  {Rioja} M.,   {Dodson} R.,  2012, \mn@doi [\apj] {10.1088/0004-637X/747/1/63}, \href {https://ui.adsabs.harvard.edu/abs/2012ApJ...747...63A} {747, 63}

\bibitem[\protect\citeauthoryear{{Ahnen} et~al.,}{{Ahnen} et~al.}{2016}]{Ahnen2016}
{Ahnen} M.~L.,  et~al., 2016, \mn@doi [\aap] {10.1051/0004-6361/201628447}, \href {https://ui.adsabs.harvard.edu/abs/2016A&A...593A..91A} {593, A91}

\bibitem[\protect\citeauthoryear{{Ajello} et~al.,}{{Ajello} et~al.}{2022}]{Ajello2022}
{Ajello} M.,  et~al., 2022, \mn@doi [\apjs] {10.3847/1538-4365/ac9523}, \href {https://ui.adsabs.harvard.edu/abs/2022ApJS..263...24A} {263, 24}

\bibitem[\protect\citeauthoryear{{Albert} et~al.,}{{Albert} et~al.}{2006}]{Albert2006}
{Albert} J.,  et~al., 2006, \mn@doi [\apjl] {10.1086/504845}, \href {https://ui.adsabs.harvard.edu/abs/2006ApJ...642L.119A} {642, L119}

\bibitem[\protect\citeauthoryear{{Aleksi{\'c}} et~al.,}{{Aleksi{\'c}} et~al.}{2014}]{Aleksic2014}
{Aleksi{\'c}} J.,  et~al., 2014, \mn@doi [\aap] {10.1051/0004-6361/201423364}, \href {https://ui.adsabs.harvard.edu/abs/2014A&A...567A.135A} {567, A135}

\bibitem[\protect\citeauthoryear{{Aller}, {Aller}, {Hughes}  \& {Latimer}}{{Aller} et~al.}{1999}]{Aller1999}
{Aller} M.~F.,  {Aller} H.~D.,  {Hughes} P.~A.,   {Latimer} G.~E.,  1999, \mn@doi [\apj] {10.1086/306799}, \href {https://ui.adsabs.harvard.edu/abs/1999ApJ...512..601A} {512, 601}

\bibitem[\protect\citeauthoryear{{Angel} \& {Stockman}}{{Angel} \& {Stockman}}{1980}]{Angel1980}
{Angel} J.~R.~P.,  {Stockman} H.~S.,  1980, \mn@doi [\araa] {10.1146/annurev.aa.18.090180.001541}, \href {https://ui.adsabs.harvard.edu/abs/1980ARA&A..18..321A} {18, 321}

\bibitem[\protect\citeauthoryear{{Astropy Collaboration} et~al.,}{{Astropy Collaboration} et~al.}{2013}]{astropy:2013}
{Astropy Collaboration} et~al., 2013, \mn@doi [\aap] {10.1051/0004-6361/201322068}, \href {http://adsabs.harvard.edu/abs/2013A%26A...558A..33A} {558, A33}

\bibitem[\protect\citeauthoryear{{Astropy Collaboration} et~al.,}{{Astropy Collaboration} et~al.}{2018}]{astropy:2018}
{Astropy Collaboration} et~al., 2018, \mn@doi [\aj] {10.3847/1538-3881/aabc4f}, \href {https://ui.adsabs.harvard.edu/abs/2018AJ....156..123A} {156, 123}

\bibitem[\protect\citeauthoryear{{Baghel}, {Kharb}, {Silpa}, {Ho}  \& {Harrison}}{{Baghel} et~al.}{2023}]{Baghel2023}
{Baghel} J.,  {Kharb} P.,  {Silpa} {Ho} L.~C.,   {Harrison} C.~M.,  2023, \mn@doi [\mnras] {10.1093/mnras/stac3691}, \href {https://ui.adsabs.harvard.edu/abs/2023MNRAS.519.2773B} {519, 2773}

\bibitem[\protect\citeauthoryear{{Baldi} \& {Capetti}}{{Baldi} \& {Capetti}}{2009}]{Baldi2009}
{Baldi} R.~D.,  {Capetti} A.,  2009, \mn@doi [\aap] {10.1051/0004-6361/200913021}, \href {https://ui.adsabs.harvard.edu/abs/2009A&A...508..603B} {508, 603}

\bibitem[\protect\citeauthoryear{{Baldi}, {Capetti}  \& {Giovannini}}{{Baldi} et~al.}{2016}]{Baldi2016}
{Baldi} R.~D.,  {Capetti} A.,   {Giovannini} G.,  2016, \mn@doi [Astronomische Nachrichten] {10.1002/asna.201512275}, \href {https://ui.adsabs.harvard.edu/abs/2016AN....337..114B} {337, 114}

\bibitem[\protect\citeauthoryear{{Bauer}, {Condon}, {Thuan}  \& {Broderick}}{{Bauer} et~al.}{2000}]{Bauer2000}
{Bauer} F.~E.,  {Condon} J.~J.,  {Thuan} T.~X.,   {Broderick} J.~J.,  2000, \mn@doi [\apjs] {10.1086/313425}, \href {https://ui.adsabs.harvard.edu/abs/2000ApJS..129..547B} {129, 547}

\bibitem[\protect\citeauthoryear{{Becker}, {White}  \& {Helfand}}{{Becker} et~al.}{1995}]{Becker1995}
{Becker} R.~H.,  {White} R.~L.,   {Helfand} D.~J.,  1995, \mn@doi [\apj] {10.1086/176166}, \href {https://ui.adsabs.harvard.edu/abs/1995ApJ...450..559B} {450, 559}

\bibitem[\protect\citeauthoryear{{Bhattacharya}, {Sreekumar}, {Mukhopadhyay}  \& {Tomar}}{{Bhattacharya} et~al.}{2016}]{Bhattacharya2016}
{Bhattacharya} D.,  {Sreekumar} P.,  {Mukhopadhyay} B.,   {Tomar} I.,  2016, \mn@doi [Research in Astronomy and Astrophysics] {10.1088/1674-4527/16/4/054}, \href {https://ui.adsabs.harvard.edu/abs/2016RAA....16...54B} {16, 54}

\bibitem[\protect\citeauthoryear{{Blandford} \& {K{\"o}nigl}}{{Blandford} \& {K{\"o}nigl}}{1979}]{Blandford1979}
{Blandford} R.~D.,  {K{\"o}nigl} A.,  1979, \mn@doi [\apj] {10.1086/157262}, \href {https://ui.adsabs.harvard.edu/abs/1979ApJ...232...34B} {232, 34}

\bibitem[\protect\citeauthoryear{{Blandford} \& {Payne}}{{Blandford} \& {Payne}}{1982}]{BP1982}
{Blandford} R.~D.,  {Payne} D.~G.,  1982, \mnras, \href {https://ui.adsabs.harvard.edu/abs/1982MNRAS.199..883B} {199, 883}

\bibitem[\protect\citeauthoryear{{Blandford} \& {Rees}}{{Blandford} \& {Rees}}{1978}]{Blandford1978}
{Blandford} R.~D.,  {Rees} M.~J.,  1978, \mn@doi [\physscr] {10.1088/0031-8949/17/3/020}, \href {https://ui.adsabs.harvard.edu/abs/1978PhyS...17..265B} {17, 265}

\bibitem[\protect\citeauthoryear{{Blandford} \& {Znajek}}{{Blandford} \& {Znajek}}{1977}]{BZ1977}
{Blandford} R.~D.,  {Znajek} R.~L.,  1977, \mnras, \href {https://ui.adsabs.harvard.edu/abs/1977MNRAS.179..433B} {179, 433}

\bibitem[\protect\citeauthoryear{{Blundell} \& {Rawlings}}{{Blundell} \& {Rawlings}}{2000}]{Blundell2000}
{Blundell} K.~M.,  {Rawlings} S.,  2000, \mn@doi [\aj] {10.1086/301254}, \href {https://ui.adsabs.harvard.edu/abs/2000AJ....119.1111B} {119, 1111}

\bibitem[\protect\citeauthoryear{{Bridle}}{{Bridle}}{1982}]{Bridle1982}
{Bridle} A.~H.,  1982, in {Heeschen} D.~S.,  {Wade} C.~M.,  eds, ~ Vol. 97, Extragalactic Radio Sources. pp 121--128

\bibitem[\protect\citeauthoryear{{Bridle} \& {Perley}}{{Bridle} \& {Perley}}{1984}]{Bridle1984}
{Bridle} A.~H.,  {Perley} R.~A.,  1984, \mn@doi [\araa] {10.1146/annurev.aa.22.090184.001535}, \href {https://ui.adsabs.harvard.edu/abs/1984ARA&A..22..319B} {22, 319}

\bibitem[\protect\citeauthoryear{{Bridle}, {Hough}, {Lonsdale}, {Burns}  \& {Laing}}{{Bridle} et~al.}{1994}]{Bridle1994}
{Bridle} A.~H.,  {Hough} D.~H.,  {Lonsdale} C.~J.,  {Burns} J.~O.,   {Laing} R.~A.,  1994, \mn@doi [\aj] {10.1086/117112}, \href {https://ui.adsabs.harvard.edu/abs/1994AJ....108..766B} {108, 766}

\bibitem[\protect\citeauthoryear{{Butuzova}}{{Butuzova}}{2021}]{Butuzova2021}
{Butuzova} M.~S.,  2021, \mn@doi [Astronomy Reports] {10.1134/S1063772921080023}, \href {https://ui.adsabs.harvard.edu/abs/2021ARep...65..635B} {65, 635}

\bibitem[\protect\citeauthoryear{{CASA Team} et~al.,}{{CASA Team} et~al.}{2022}]{CASA2022}
{CASA Team} et~al., 2022, \mn@doi [\pasp] {10.1088/1538-3873/ac9642}, \href {https://ui.adsabs.harvard.edu/abs/2022PASP..134k4501C} {134, 114501}

\bibitem[\protect\citeauthoryear{{Cassaro}, {Stanghellini}, {Bondi}, {Dallacasa}, {della Ceca}  \& {Zappal{\`a}}}{{Cassaro} et~al.}{1999}]{Cassaro1999}
{Cassaro} P.,  {Stanghellini} C.,  {Bondi} M.,  {Dallacasa} D.,  {della Ceca} R.,   {Zappal{\`a}} R.~A.,  1999, \mn@doi [\aaps] {10.1051/aas:1999511}, \href {https://ui.adsabs.harvard.edu/abs/1999A&AS..139..601C} {139, 601}

\bibitem[\protect\citeauthoryear{{Cassaro}, {Stanghellini}, {Dallacasa}, {Bondi}  \& {Zappal{\`a}}}{{Cassaro} et~al.}{2002}]{Cassaro2002}
{Cassaro} P.,  {Stanghellini} C.,  {Dallacasa} D.,  {Bondi} M.,   {Zappal{\`a}} R.~A.,  2002, \mn@doi [\aap] {10.1051/0004-6361:20011460}, \href {https://ui.adsabs.harvard.edu/abs/2002A&A...381..378C} {381, 378}

\bibitem[\protect\citeauthoryear{{Cawthorne} \& {Hughes}}{{Cawthorne} \& {Hughes}}{2013}]{Cawthrone2013}
{Cawthorne} T.~V.,  {Hughes} P.~A.,  2013, \mn@doi [\apj] {10.1088/0004-637X/771/1/60}, \href {https://ui.adsabs.harvard.edu/abs/2013ApJ...771...60C} {771, 60}

\bibitem[\protect\citeauthoryear{{Cawthorne}, {Wardle}, {Roberts}  \& {Gabuzda}}{{Cawthorne} et~al.}{1993}]{Cawthrone1993}
{Cawthorne} T.~V.,  {Wardle} J.~F.~C.,  {Roberts} D.~H.,   {Gabuzda} D.~C.,  1993, \apj, \href {https://ui.adsabs.harvard.edu/abs/1993ApJ...416..519C} {416, 519}

\bibitem[\protect\citeauthoryear{{Cerruti}, {Benbow}, {Chen}, {Dumm}, {Fortson}  \& {Shahinyan}}{{Cerruti} et~al.}{2017}]{Cerruti2017}
{Cerruti} M.,  {Benbow} W.,  {Chen} X.,  {Dumm} J.~P.,  {Fortson} L.~F.,   {Shahinyan} K.,  2017, \mn@doi [\aap] {10.1051/0004-6361/201730799}, \href {https://ui.adsabs.harvard.edu/abs/2017A&A...606A..68C} {606, A68}

\bibitem[\protect\citeauthoryear{{Cheng} \& {An}}{{Cheng} \& {An}}{2018}]{Cheng2018}
{Cheng} X.~P.,  {An} T.,  2018, \mn@doi [\apj] {10.3847/1538-4357/aad22c}, \href {https://ui.adsabs.harvard.edu/abs/2018ApJ...863..155C} {863, 155}

\bibitem[\protect\citeauthoryear{{Cooke} et~al.,}{{Cooke} et~al.}{1978}]{Cooke1978}
{Cooke} B.~A.,  et~al., 1978, \mn@doi [\mnras] {10.1093/mnras/182.3.489}, \href {https://ui.adsabs.harvard.edu/abs/1978MNRAS.182..489C} {182, 489}

\bibitem[\protect\citeauthoryear{{Das}, {Kharb}, {Morganti}  \& {Nandi}}{{Das} et~al.}{2021}]{Das2021}
{Das} S.,  {Kharb} P.,  {Morganti} R.,   {Nandi} S.,  2021, \mn@doi [\mnras] {10.1093/mnras/stab1148}, \href {https://ui.adsabs.harvard.edu/abs/2021MNRAS.504.4416D} {504, 4416}

\bibitem[\protect\citeauthoryear{{Di Gesu} et~al.,}{{Di Gesu} et~al.}{2022}]{Laura2022}
{Di Gesu} L.,  et~al., 2022, \mn@doi [\apjl] {10.3847/2041-8213/ac913a}, \href {https://ui.adsabs.harvard.edu/abs/2022ApJ...938L...7D} {938, L7}

\bibitem[\protect\citeauthoryear{{Fan} et~al.,}{{Fan} et~al.}{2006}]{Fan2006}
{Fan} J.-H.,  et~al., 2006, \mn@doi [\pasj] {10.1093/pasj/58.6.945}, \href {https://ui.adsabs.harvard.edu/abs/2006PASJ...58..945F} {58, 945}

\bibitem[\protect\citeauthoryear{{Fanaroff} \& {Riley}}{{Fanaroff} \& {Riley}}{1974}]{Fanaroff1974}
{Fanaroff} B.~L.,  {Riley} J.~M.,  1974, \mnras, \href {https://ui.adsabs.harvard.edu/abs/1974MNRAS.167P..31F} {167, 31P}

\bibitem[\protect\citeauthoryear{{Farnes}, {Green}  \& {Kantharia}}{{Farnes} et~al.}{2014}]{Farnes2014}
{Farnes} J.~S.,  {Green} D.~A.,   {Kantharia} N.~G.,  2014, \mn@doi [\mnras] {10.1093/mnras/stt2118}, \href {https://ui.adsabs.harvard.edu/abs/2014MNRAS.437.3236F} {437, 3236}

\bibitem[\protect\citeauthoryear{{Fey} \& {Charlot}}{{Fey} \& {Charlot}}{1997}]{Fey1997}
{Fey} A.~L.,  {Charlot} P.,  1997, \mn@doi [\apjs] {10.1086/313017}, \href {https://ui.adsabs.harvard.edu/abs/1997ApJS..111...95F} {111, 95}

\bibitem[\protect\citeauthoryear{{Fey} \& {Charlot}}{{Fey} \& {Charlot}}{2000}]{Fey2000}
{Fey} A.~L.,  {Charlot} P.,  2000, \mn@doi [\apjs] {10.1086/313382}, \href {https://ui.adsabs.harvard.edu/abs/2000ApJS..128...17F} {128, 17}

\bibitem[\protect\citeauthoryear{Fleming, Green, Jannuzi, Liebert, Smith  \& Fink}{Fleming et~al.}{1993}]{Fleming1993}
Fleming T.~A.,  Green R.~F.,  Jannuzi B.~T.,  Liebert J.,  Smith P.~S.,   Fink H.,  1993, \aj, 106, 1729

\bibitem[\protect\citeauthoryear{Fossati, Maraschi, Celotti, Comastri  \& Ghisellini}{Fossati et~al.}{1998}]{Fossati1998}
Fossati G.,  Maraschi L.,  Celotti A.,  Comastri A.,   Ghisellini G.,  1998, \mnras, 299, 433–448

\bibitem[\protect\citeauthoryear{{Gabuzda}, {Wardle}  \& {Roberts}}{{Gabuzda} et~al.}{1989}]{1989Gabuzda}
{Gabuzda} D.~C.,  {Wardle} J. F.~C.,   {Roberts} D.~H.,  1989, \mn@doi [\apjl] {10.1086/185361}, \href {https://ui.adsabs.harvard.edu/abs/1989ApJ...336L..59G} {336, L59}

\bibitem[\protect\citeauthoryear{{Gabuzda}, {Cawthorne}, {Roberts}  \& {Wardle}}{{Gabuzda} et~al.}{1992}]{Gabuzda1992}
{Gabuzda} D.~C.,  {Cawthorne} T.~V.,  {Roberts} D.~H.,   {Wardle} J.~F.~C.,  1992, \mn@doi [\apj] {10.1086/171128}, \href {https://ui.adsabs.harvard.edu/abs/1992ApJ...388...40G} {388, 40}

\bibitem[\protect\citeauthoryear{{Gabuzda}, {Sitko}  \& {Smith}}{{Gabuzda} et~al.}{1996}]{Gabuzda1996}
{Gabuzda} D.~C.,  {Sitko} M.~L.,   {Smith} P.~S.,  1996, \mn@doi [\aj] {10.1086/118149}, \href {https://ui.adsabs.harvard.edu/abs/1996AJ....112.1877G} {112, 1877}

\bibitem[\protect\citeauthoryear{{Gallimore}, {Axon}, {O'Dea}, {Baum}  \& {Pedlar}}{{Gallimore} et~al.}{2006}]{Gallimore2006}
{Gallimore} J.~F.,  {Axon} D.~J.,  {O'Dea} C.~P.,  {Baum} S.~A.,   {Pedlar} A.,  2006, \mn@doi [\aj] {10.1086/504593}, \href {https://ui.adsabs.harvard.edu/abs/2006AJ....132..546G} {132, 546}

\bibitem[\protect\citeauthoryear{{Ghisellini}}{{Ghisellini}}{2011}]{Ghisellini2011b}
{Ghisellini} G.,  2011, in {Aharonian} F.~A.,  {Hofmann} W.,   {Rieger} F.~M.,  eds,  American Institute of Physics Conference Series Vol. 1381, 25th Texas Symposium on Relativistic AstroPhysics (Texas 2010). pp 180--198 (\mn@eprint {arXiv} {1104.0006}), \mn@doi{10.1063/1.3635832}

\bibitem[\protect\citeauthoryear{{Ghisellini}, {Celotti}, {Fossati}, {Maraschi}  \& {Comastri}}{{Ghisellini} et~al.}{1998}]{Ghisellini1998}
{Ghisellini} G.,  {Celotti} A.,  {Fossati} G.,  {Maraschi} L.,   {Comastri} A.,  1998, \mn@doi [\mnras] {10.1046/j.1365-8711.1998.02032.x}, \href {https://ui.adsabs.harvard.edu/abs/1998MNRAS.301..451G} {301, 451}

\bibitem[\protect\citeauthoryear{{Ghisellini}, {Tavecchio}, {Foschini}, {Ghirlanda}, {Maraschi}  \& {Celotti}}{{Ghisellini} et~al.}{2010}]{Ghisellini2010}
{Ghisellini} G.,  {Tavecchio} F.,  {Foschini} L.,  {Ghirlanda} G.,  {Maraschi} L.,   {Celotti} A.,  2010, \mn@doi [\mnras] {10.1111/j.1365-2966.2009.15898.x}, \href {https://ui.adsabs.harvard.edu/abs/2010MNRAS.402..497G} {402, 497}

\bibitem[\protect\citeauthoryear{{Ghisellini}, {Tavecchio}, {Foschini}, {Bonnoli}  \& {Tagliaferri}}{{Ghisellini} et~al.}{2013}]{Ghisellini2013}
{Ghisellini} G.,  {Tavecchio} F.,  {Foschini} L.,  {Bonnoli} G.,   {Tagliaferri} G.,  2013, \mn@doi [\mnras] {10.1093/mnrasl/slt041}, \href {https://ui.adsabs.harvard.edu/abs/2013MNRAS.432L..66G} {432, L66}

\bibitem[\protect\citeauthoryear{{Ghisellini}, {Righi}, {Costamante}  \& {Tavecchio}}{{Ghisellini} et~al.}{2017}]{Ghisellini2017}
{Ghisellini} G.,  {Righi} C.,  {Costamante} L.,   {Tavecchio} F.,  2017, \mn@doi [\mnras] {10.1093/mnras/stx806}, \href {https://ui.adsabs.harvard.edu/abs/2017MNRAS.469..255G} {469, 255}

\bibitem[\protect\citeauthoryear{{Giroletti}, {Giovannini}, {Taylor}  \& {Falomo}}{{Giroletti} et~al.}{2004}]{Giroletti2004}
{Giroletti} M.,  {Giovannini} G.,  {Taylor} G.~B.,   {Falomo} R.,  2004, \mn@doi [\apj] {10.1086/423231}, \href {https://ui.adsabs.harvard.edu/abs/2004ApJ...613..752G} {613, 752}

\bibitem[\protect\citeauthoryear{{Giroletti}, {Giovannini}, {Taylor}  \& {Falomo}}{{Giroletti} et~al.}{2006}]{Giroletti2006}
{Giroletti} M.,  {Giovannini} G.,  {Taylor} G.~B.,   {Falomo} R.,  2006, \mn@doi [\apj] {10.1086/504971}, \href {https://ui.adsabs.harvard.edu/abs/2006ApJ...646..801G} {646, 801}

\bibitem[\protect\citeauthoryear{{G{\'o}mez} et~al.,}{{G{\'o}mez} et~al.}{2022}]{Gomez2022}
{G{\'o}mez} J.~L.,  et~al., 2022, \mn@doi [\apj] {10.3847/1538-4357/ac3bcc}, \href {https://ui.adsabs.harvard.edu/abs/2022ApJ...924..122G} {924, 122}

\bibitem[\protect\citeauthoryear{{Green}, {Schmidt}  \& {Liebert}}{{Green} et~al.}{1986}]{Green1986}
{Green} R.~F.,  {Schmidt} M.,   {Liebert} J.,  1986, \apjs, \href {https://ui.adsabs.harvard.edu/abs/1986ApJS...61..305G} {61, 305}

\bibitem[\protect\citeauthoryear{{Hardcastle} \& {Krause}}{{Hardcastle} \& {Krause}}{2014}]{Hardcastle2014}
{Hardcastle} M.~J.,  {Krause} M.~G.~H.,  2014, \mn@doi [\mnras] {10.1093/mnras/stu1229}, \href {https://ui.adsabs.harvard.edu/abs/2014MNRAS.443.1482H} {443, 1482}

\bibitem[\protect\citeauthoryear{{Hawley}, {Fendt}, {Hardcastle}, {Nokhrina}  \& {Tchekhovskoy}}{{Hawley} et~al.}{2015}]{Hawley2015}
{Hawley} J.~F.,  {Fendt} C.,  {Hardcastle} M.,  {Nokhrina} E.,   {Tchekhovskoy} A.,  2015, \mn@doi [\ssr] {10.1007/s11214-015-0174-7}, \href {https://ui.adsabs.harvard.edu/abs/2015SSRv..191..441H} {191, 441}

\bibitem[\protect\citeauthoryear{{Heidt} \& {Wagner}}{{Heidt} \& {Wagner}}{1996}]{Heidt1996}
{Heidt} J.,  {Wagner} S.~J.,  1996, \mn@doi [\aap] {10.48550/arXiv.astro-ph/9506032}, \href {https://ui.adsabs.harvard.edu/abs/1996A&A...305...42H} {305, 42}

\bibitem[\protect\citeauthoryear{{Helmboldt} et~al.,}{{Helmboldt} et~al.}{2007}]{Helmboldt2007}
{Helmboldt} J.~F.,  et~al., 2007, \mn@doi [\apj] {10.1086/511005}, \href {https://ui.adsabs.harvard.edu/abs/2007ApJ...658..203H} {658, 203}

\bibitem[\protect\citeauthoryear{{Hovatta} \& {Lindfors}}{{Hovatta} \& {Lindfors}}{2019}]{Hovatta2019}
{Hovatta} T.,  {Lindfors} E.,  2019, \mn@doi [\nar] {10.1016/j.newar.2020.101541}, \href {https://ui.adsabs.harvard.edu/abs/2019NewAR..8701541H} {87, 101541}

\bibitem[\protect\citeauthoryear{{Hovatta}, {Lister}, {Aller}, {Aller}, {Homan}, {Kovalev}, {Pushkarev}  \& {Savolainen}}{{Hovatta} et~al.}{2012}]{Hovatta2012}
{Hovatta} T.,  {Lister} M.~L.,  {Aller} M.~F.,  {Aller} H.~D.,  {Homan} D.~C.,  {Kovalev} Y.~Y.,  {Pushkarev} A.~B.,   {Savolainen} T.,  2012, \mn@doi [\aj] {10.1088/0004-6256/144/4/105}, \href {https://ui.adsabs.harvard.edu/abs/2012AJ....144..105H} {144, 105}

\bibitem[\protect\citeauthoryear{{Hovatta} et~al.,}{{Hovatta} et~al.}{2016}]{Hovatta2016}
{Hovatta} T.,  et~al., 2016, \mn@doi [\aap] {10.1051/0004-6361/201628974}, \href {https://ui.adsabs.harvard.edu/abs/2016A&A...596A..78H} {596, A78}

\bibitem[\protect\citeauthoryear{{Impey} \& {Tapia}}{{Impey} \& {Tapia}}{1988}]{Impey1988}
{Impey} C.~D.,  {Tapia} S.,  1988, \mn@doi [\apj] {10.1086/166775}, \href {https://ui.adsabs.harvard.edu/abs/1988ApJ...333..666I} {333, 666}

\bibitem[\protect\citeauthoryear{{Intema}, {Jagannathan}, {Mooley}  \& {Frail}}{{Intema} et~al.}{2017}]{Intema2017}
{Intema} H.~T.,  {Jagannathan} P.,  {Mooley} K.~P.,   {Frail} D.~A.,  2017, \mn@doi [\aap] {10.1051/0004-6361/201628536}, \href {https://ui.adsabs.harvard.edu/abs/2017A&A...598A..78I} {598, A78}

\bibitem[\protect\citeauthoryear{{Jannuzi}, {Smith}  \& {Elston}}{{Jannuzi} et~al.}{1993}]{Jannuzi1993}
{Jannuzi} B.~T.,  {Smith} P.~S.,   {Elston} R.,  1993, \mn@doi [\apjs] {10.1086/191765}, \href {https://ui.adsabs.harvard.edu/abs/1993ApJS...85..265J} {85, 265}

\bibitem[\protect\citeauthoryear{{Jorstad} et~al.,}{{Jorstad} et~al.}{2017}]{Jorstad2017}
{Jorstad} S.~G.,  et~al., 2017, \mn@doi [\apj] {10.3847/1538-4357/aa8407}, \href {https://ui.adsabs.harvard.edu/abs/2017ApJ...846...98J} {846, 98}

\bibitem[\protect\citeauthoryear{{Keenan}, {Meyer}, {Georganopoulos}, {Reddy}  \& {French}}{{Keenan} et~al.}{2021}]{Keenan2021}
{Keenan} M.,  {Meyer} E.~T.,  {Georganopoulos} M.,  {Reddy} K.,   {French} O.~J.,  2021, \mn@doi [\mnras] {10.1093/mnras/stab1182}, \href {https://ui.adsabs.harvard.edu/abs/2021MNRAS.505.4726K} {505, 4726}

\bibitem[\protect\citeauthoryear{{Kellermann}, {Sramek}, {Schmidt}, {Shaffer}  \& {Green}}{{Kellermann} et~al.}{1989}]{Kellermann1989}
{Kellermann} K.~I.,  {Sramek} R.,  {Schmidt} M.,  {Shaffer} D.~B.,   {Green} R.,  1989, \aj, \href {https://ui.adsabs.harvard.edu/abs/1989AJ.....98.1195K} {98, 1195}

\bibitem[\protect\citeauthoryear{{Kharb} \& {Shastri}}{{Kharb} \& {Shastri}}{2004}]{Kharb2004}
{Kharb} P.,  {Shastri} P.,  2004, \mn@doi [\aap] {10.1051/0004-6361:200400023}, \href {https://ui.adsabs.harvard.edu/abs/2004A&A...425..825K} {425, 825}

\bibitem[\protect\citeauthoryear{{Kharb}, {O'Dea}, {Baum}, {Daly}, {Mory}, {Donahue}  \& {Guerra}}{{Kharb} et~al.}{2008a}]{Kharb2008b}
{Kharb} P.,  {O'Dea} C.~P.,  {Baum} S.~A.,  {Daly} R.~A.,  {Mory} M.~P.,  {Donahue} M.,   {Guerra} E.~J.,  2008a, \mn@doi [\apjs] {10.1086/520840}, \href {https://ui.adsabs.harvard.edu/abs/2008ApJS..174...74K} {174, 74}

\bibitem[\protect\citeauthoryear{{Kharb}, {Gabuzda}  \& {Shastri}}{{Kharb} et~al.}{2008b}]{Kharb2008}
{Kharb} P.,  {Gabuzda} D.,   {Shastri} P.,  2008b, \mn@doi [\mnras] {10.1111/j.1365-2966.2007.12690.x}, \href {https://ui.adsabs.harvard.edu/abs/2008MNRAS.384..230K} {384, 230}

\bibitem[\protect\citeauthoryear{{Kharb}, {Lister}  \& {Cooper}}{{Kharb} et~al.}{2010}]{Kharb10}
{Kharb} P.,  {Lister} M.~L.,   {Cooper} N.~J.,  2010, \apj, 710, 764

\bibitem[\protect\citeauthoryear{{Kharb}, {Sasikumar}, {Baghel}  \& {Ghosh}}{{Kharb} et~al.}{2023}]{Kharb23}
{Kharb} P.,  {Sasikumar} S.,  {Baghel} J.,   {Ghosh} S.,  2023, \mn@doi [arXiv e-prints] {10.48550/arXiv.2305.04420}, \href {https://ui.adsabs.harvard.edu/abs/2023arXiv230504420K} {p. arXiv:2305.04420}

\bibitem[\protect\citeauthoryear{{Kinman}}{{Kinman}}{1975}]{Kinman1975}
{Kinman} T.~D.,  1975, in {Sherwood} V.~E.,  {Plaut} L.,  eds, ~ Vol. 67, Variable Stars and Stellar Evolution. p.~573

\bibitem[\protect\citeauthoryear{{Kormendy} \& {Ho}}{{Kormendy} \& {Ho}}{2013}]{Kormendy2013}
{Kormendy} J.,  {Ho} L.~C.,  2013, \mn@doi [\araa] {10.1146/annurev-astro-082708-101811}, \href {https://ui.adsabs.harvard.edu/abs/2013ARA&A..51..511K} {51, 511}

\bibitem[\protect\citeauthoryear{{Kuehr}, {Witzel}, {Pauliny-Toth}  \& {Nauber}}{{Kuehr} et~al.}{1981}]{Kuehr1981}
{Kuehr} H.,  {Witzel} A.,  {Pauliny-Toth} I.~I.~K.,   {Nauber} U.,  1981, \aaps, \href {https://ui.adsabs.harvard.edu/abs/1981A&AS...45..367K} {45, 367}

\bibitem[\protect\citeauthoryear{{Laing} \& {Bridle}}{{Laing} \& {Bridle}}{2013}]{Laing2013}
{Laing} R.~A.,  {Bridle} A.~H.,  2013, \mn@doi [\mnras] {10.1093/mnras/stt531}, \href {https://ui.adsabs.harvard.edu/abs/2013MNRAS.432.1114L} {432, 1114}

\bibitem[\protect\citeauthoryear{{Laing} \& {Bridle}}{{Laing} \& {Bridle}}{2014}]{Laing2014}
{Laing} R.~A.,  {Bridle} A.~H.,  2014, \mn@doi [\mnras] {10.1093/mnras/stt2138}, \href {https://ui.adsabs.harvard.edu/abs/2014MNRAS.437.3405L} {437, 3405}

\bibitem[\protect\citeauthoryear{{Landt}, {Perlman}  \& {Padovani}}{{Landt} et~al.}{2006}]{Landt2006}
{Landt} H.,  {Perlman} E.~S.,   {Padovani} P.,  2006, \apj, \href {https://ui.adsabs.harvard.edu/abs/2006ApJ...637..183L} {637, 183}

\bibitem[\protect\citeauthoryear{{Laurent-Muehleisen}, {Kollgaard}, {Moellenbrock}  \& {Feigelson}}{{Laurent-Muehleisen} et~al.}{1993}]{Laurent1993}
{Laurent-Muehleisen} S.~A.,  {Kollgaard} R.~I.,  {Moellenbrock} G.~A.,   {Feigelson} E.~D.,  1993, \mn@doi [\aj] {10.1086/116691}, \href {https://ui.adsabs.harvard.edu/abs/1993AJ....106..875L} {106, 875}

\bibitem[\protect\citeauthoryear{{Laurent-Muehleisen}, {Kollgaard}, {Feigelson}, {Brinkmann}  \& {Siebert}}{{Laurent-Muehleisen} et~al.}{1999}]{Laurent1999}
{Laurent-Muehleisen} S.~A.,  {Kollgaard} R.~I.,  {Feigelson} E.~D.,  {Brinkmann} W.,   {Siebert} J.,  1999, \mn@doi [\apj] {10.1086/307881}, \href {https://ui.adsabs.harvard.edu/abs/1999ApJ...525..127L} {525, 127}

\bibitem[\protect\citeauthoryear{{Lico} et~al.,}{{Lico} et~al.}{2020}]{Lico2020}
{Lico} R.,  et~al., 2020, \mn@doi [\aap] {10.1051/0004-6361/201936564}, \href {https://ui.adsabs.harvard.edu/abs/2020A&A...634A..87L} {634, A87}

\bibitem[\protect\citeauthoryear{{Lister} \& {Homan}}{{Lister} \& {Homan}}{2005}]{Lister2005}
{Lister} M.~L.,  {Homan} D.~C.,  2005, \mn@doi [\aj] {10.1086/432969}, \href {https://ui.adsabs.harvard.edu/abs/2005AJ....130.1389L} {130, 1389}

\bibitem[\protect\citeauthoryear{{Lister} et~al.,}{{Lister} et~al.}{2011}]{Lister2011}
{Lister} M.~L.,  et~al., 2011, \mn@doi [\apj] {10.1088/0004-637X/742/1/27}, \href {https://ui.adsabs.harvard.edu/abs/2011ApJ...742...27L} {742, 27}

\bibitem[\protect\citeauthoryear{Lister et~al.,}{Lister et~al.}{2013}]{Lister2013}
Lister M.~L.,  et~al., 2013, \aj, \href {https://ui.adsabs.harvard.edu/abs/2013AJ....146..120L} {146, 120}

\bibitem[\protect\citeauthoryear{{Lister}, {Aller}, {Aller}, {Hovatta}, {Max-Moerbeck}, {Readhead}, {Richards}  \& {Ros}}{{Lister} et~al.}{2015}]{Lister2015}
{Lister} M.~L.,  {Aller} M.~F.,  {Aller} H.~D.,  {Hovatta} T.,  {Max-Moerbeck} W.,  {Readhead} A.~C.~S.,  {Richards} J.~L.,   {Ros} E.,  2015, \mn@doi [\apjl] {10.1088/2041-8205/810/1/L9}, \href {https://ui.adsabs.harvard.edu/abs/2015ApJ...810L...9L} {810, L9}

\bibitem[\protect\citeauthoryear{{Lister} et~al.,}{{Lister} et~al.}{2016}]{Lister2016}
{Lister} M.~L.,  et~al., 2016, \mn@doi [\aj] {10.3847/0004-6256/152/1/12}, \href {https://ui.adsabs.harvard.edu/abs/2016AJ....152...12L} {152, 12}

\bibitem[\protect\citeauthoryear{{Lister}, {Aller}, {Aller}, {Hodge}, {Homan}, {Kovalev}, {Pushkarev}  \& {Savolainen}}{{Lister} et~al.}{2018}]{Lister2018}
{Lister} M.~L.,  {Aller} M.~F.,  {Aller} H.~D.,  {Hodge} M.~A.,  {Homan} D.~C.,  {Kovalev} Y.~Y.,  {Pushkarev} A.~B.,   {Savolainen} T.,  2018, \mn@doi [\apjs] {10.3847/1538-4365/aa9c44}, \href {https://ui.adsabs.harvard.edu/abs/2018ApJS..234...12L} {234, 12}

\bibitem[\protect\citeauthoryear{{Lister} et~al.,}{{Lister} et~al.}{2019}]{Lister2019}
{Lister} M.~L.,  et~al., 2019, \mn@doi [\apj] {10.3847/1538-4357/ab08ee}, \href {https://ui.adsabs.harvard.edu/abs/2019ApJ...874...43L} {874, 43}

\bibitem[\protect\citeauthoryear{{Lister}, {Homan}, {Kellermann}, {Kovalev}, {Pushkarev}, {Ros}  \& {Savolainen}}{{Lister} et~al.}{2021}]{Lister2021}
{Lister} M.~L.,  {Homan} D.~C.,  {Kellermann} K.~I.,  {Kovalev} Y.~Y.,  {Pushkarev} A.~B.,  {Ros} E.,   {Savolainen} T.,  2021, \mn@doi [\apj] {10.3847/1538-4357/ac230f}, \href {https://ui.adsabs.harvard.edu/abs/2021ApJ...923...30L} {923, 30}

\bibitem[\protect\citeauthoryear{{Lyutikov}, {Pariev}  \& {Gabuzda}}{{Lyutikov} et~al.}{2005}]{Lyutikov2005}
{Lyutikov} M.,  {Pariev} V.,   {Gabuzda} D.,  2005, Mem. S.A.It, 76, 114

\bibitem[\protect\citeauthoryear{{Machalski} \& {Condon}}{{Machalski} \& {Condon}}{1985}]{Machalski1985}
{Machalski} J.,  {Condon} J.~J.,  1985, \mn@doi [\aj] {10.1086/113703}, \href {https://ui.adsabs.harvard.edu/abs/1985AJ.....90....5M} {90, 5}

\bibitem[\protect\citeauthoryear{{Madejski} \& {Sikora}}{{Madejski} \& {Sikora}}{2016}]{Madejski2016}
{Madejski} G.~G.,  {Sikora} M.,  2016, \mn@doi [\araa] {10.1146/annurev-astro-081913-040044}, \href {https://ui.adsabs.harvard.edu/abs/2016ARA&A..54..725M} {54, 725}

\bibitem[\protect\citeauthoryear{{Mahatma}, {Hardcastle}, {Harwood}, {O'Sullivan}, {Heald}, {Horellou}  \& {Smith}}{{Mahatma} et~al.}{2021}]{Mahatma2021}
{Mahatma} V.~H.,  {Hardcastle} M.~J.,  {Harwood} J.,  {O'Sullivan} S.~P.,  {Heald} G.,  {Horellou} C.,   {Smith} D.~J.~B.,  2021, \mnras, \href {https://ui.adsabs.harvard.edu/abs/2021MNRAS.502..273M} {502, 273}

\bibitem[\protect\citeauthoryear{{Malik}, {Sahayanathan}, {Shah}, {Iqbal}, {Manzoor}  \& {Bhatt}}{{Malik} et~al.}{2022}]{Zahoor2022}
{Malik} Z.,  {Sahayanathan} S.,  {Shah} Z.,  {Iqbal} N.,  {Manzoor} A.,   {Bhatt} N.,  2022, \mn@doi [\mnras] {10.1093/mnras/stab3173}, \href {https://ui.adsabs.harvard.edu/abs/2022MNRAS.511..994M} {511, 994}

\bibitem[\protect\citeauthoryear{{Marchenko}}{{Marchenko}}{1985}]{Marchenko1985}
{Marchenko} S.~G.,  1985, Astrofizika, \href {https://ui.adsabs.harvard.edu/abs/1985Afz....22...15M} {22, 15}

\bibitem[\protect\citeauthoryear{{Margon}, {Jones}  \& {Wardle}}{{Margon} et~al.}{1978}]{Margon1978}
{Margon} B.,  {Jones} T.~W.,   {Wardle} J.~F.~C.,  1978, \mn@doi [\aj] {10.1086/112286}, \href {https://ui.adsabs.harvard.edu/abs/1978AJ.....83.1021M} {83, 1021}

\bibitem[\protect\citeauthoryear{{Marscher}}{{Marscher}}{2008}]{Marscher2008}
{Marscher} A.~P.,  2008, in {Rector} T.~A.,  {De Young} D.~S.,  eds,  Astronomical Society of the Pacific Conference Series Vol. 386, Extragalactic Jets: Theory and Observation from Radio to Gamma Ray. p.~437

\bibitem[\protect\citeauthoryear{{Marscher} \& {Jorstad}}{{Marscher} \& {Jorstad}}{2011}]{Marscher2011}
{Marscher} A.~P.,  {Jorstad} S.~G.,  2011, \mn@doi [\apj] {10.1088/0004-637X/729/1/26}, \href {https://ui.adsabs.harvard.edu/abs/2011ApJ...729...26M} {729, 26}

\bibitem[\protect\citeauthoryear{{Massaglia}, {Bodo}, {Rossi}, {Capetti}  \& {Mignone}}{{Massaglia} et~al.}{2016}]{Massaglia2016}
{Massaglia} S.,  {Bodo} G.,  {Rossi} P.,  {Capetti} S.,   {Mignone} A.,  2016, \mn@doi [\aap] {10.1051/0004-6361/201629375}, \href {https://ui.adsabs.harvard.edu/abs/2016A&A...596A..12M} {596, A12}

\bibitem[\protect\citeauthoryear{{Massaro}, {Capetti}, {Paggi}, {Baldi}, {Tramacere}, {Pillitteri}  \& {Campana}}{{Massaro} et~al.}{2020}]{Massaro2020}
{Massaro} F.,  {Capetti} A.,  {Paggi} A.,  {Baldi} R.~D.,  {Tramacere} A.,  {Pillitteri} I.,   {Campana} R.,  2020, \mn@doi [\apjl] {10.3847/2041-8213/abac56}, \href {https://ui.adsabs.harvard.edu/abs/2020ApJ...900L..34M} {900, L34}

\bibitem[\protect\citeauthoryear{{Mead}, {Ballard}, {Brand}, {Hough}, {Brindle}  \& {Bailey}}{{Mead} et~al.}{1990}]{Mead1990}
{Mead} A.~R.~G.,  {Ballard} K.~R.,  {Brand} P.~W.~J.~L.,  {Hough} J.~H.,  {Brindle} C.,   {Bailey} J.~A.,  1990, \aaps, \href {https://ui.adsabs.harvard.edu/abs/1990A&AS...83..183M} {83, 183}

\bibitem[\protect\citeauthoryear{{Meier}, {Koide}  \& {Uchida}}{{Meier} et~al.}{2001}]{Meier2001}
{Meier} D.~L.,  {Koide} S.,   {Uchida} Y.,  2001, \mn@doi [Science] {10.1126/science.291.5501.84}, \href {https://ui.adsabs.harvard.edu/abs/2001Sci...291...84M} {291, 84}

\bibitem[\protect\citeauthoryear{{Meyer}, {Fossati}, {Georganopoulos}  \& {Lister}}{{Meyer} et~al.}{2011}]{Meyer2011}
{Meyer} E.~T.,  {Fossati} G.,  {Georganopoulos} M.,   {Lister} M.~L.,  2011, \apj, 740, 98

\bibitem[\protect\citeauthoryear{{Meyer}, {Fossati}, {Georganopoulos}  \& {Lister}}{{Meyer} et~al.}{2012}]{Meyer2012}
{Meyer} E.~T.,  {Fossati} G.,  {Georganopoulos} M.,   {Lister} M.~L.,  2012, \mn@doi [arXiv e-prints] {10.48550/arXiv.1205.0794}, \href {https://ui.adsabs.harvard.edu/abs/2012arXiv1205.0794M} {p. arXiv:1205.0794}

\bibitem[\protect\citeauthoryear{{Miller} \& {Carini}}{{Miller} \& {Carini}}{1991}]{Miller1991}
{Miller} H.~R.,  {Carini} M.~T.,  1991, in {Miller} H.~R.,  {Wiita} P.~J.,  eds, Variability of Active Galactic Nuclei. p.~111

\bibitem[\protect\citeauthoryear{{Miller}, {Rawlings}  \& {Saunders}}{{Miller} et~al.}{1993}]{Miller1993}
{Miller} P.,  {Rawlings} S.,   {Saunders} R.,  1993, \mn@doi [\mnras] {10.1093/mnras/263.2.425}, \href {https://ui.adsabs.harvard.edu/abs/1993MNRAS.263..425M} {263, 425}

\bibitem[\protect\citeauthoryear{Mingo et~al.,}{Mingo et~al.}{2019}]{Mingo2019}
Mingo B.,  et~al., 2019, \mnras, \href {https://ui.adsabs.harvard.edu/abs/2019MNRAS.488.2701M} {488, 2701}

\bibitem[\protect\citeauthoryear{{Minns} \& {Riley}}{{Minns} \& {Riley}}{2000}]{Minns2000}
{Minns} A.~R.,  {Riley} J.~M.,  2000, \mn@doi [\mnras] {10.1046/j.1365-8711.2000.03802.x}, \href {https://ui.adsabs.harvard.edu/abs/2000MNRAS.318..827M} {318, 827}

\bibitem[\protect\citeauthoryear{{Mooney} et~al.,}{{Mooney} et~al.}{2021}]{Mooney2021}
{Mooney} S.,  et~al., 2021, \mn@doi [\apjs] {10.3847/1538-4365/ac1c0b}, \href {https://ui.adsabs.harvard.edu/abs/2021ApJS..257...30M} {257, 30}

\bibitem[\protect\citeauthoryear{{Mukherjee} \& {VERITAS Collaboration}}{{Mukherjee} \& {VERITAS Collaboration}}{2017}]{Mukherjee2017}
{Mukherjee} R.,  {VERITAS Collaboration} 2017, The Astronomer's Telegram, \href {https://ui.adsabs.harvard.edu/abs/2017ATel10051....1M} {10051, 1}

\bibitem[\protect\citeauthoryear{{Murphy}, {Browne}  \& {Perley}}{{Murphy} et~al.}{1993}]{Murphy1993}
{Murphy} D.~W.,  {Browne} I.~W.~A.,   {Perley} R.~A.,  1993, \mn@doi [\mnras] {10.1093/mnras/264.2.298}, \href {https://ui.adsabs.harvard.edu/abs/1993MNRAS.264..298M} {264, 298}

\bibitem[\protect\citeauthoryear{{Myserlis}, {Komossa}, {Angelakis}, {G{\'o}mez}, {Karamanavis}, {Krichbaum}, {Bach}  \& {Grupe}}{{Myserlis} et~al.}{2018}]{Myserlis2018}
{Myserlis} I.,  {Komossa} S.,  {Angelakis} E.,  {G{\'o}mez} J.~L.,  {Karamanavis} V.,  {Krichbaum} T.~P.,  {Bach} U.,   {Grupe} D.,  2018, \mn@doi [\aap] {10.1051/0004-6361/201732273}, \href {https://ui.adsabs.harvard.edu/abs/2018A&A...619A..88M} {619, A88}

\bibitem[\protect\citeauthoryear{{Orr} \& {Browne}}{{Orr} \& {Browne}}{1982}]{Orr1982}
{Orr} M.~J.~L.,  {Browne} I.~W.~A.,  1982, \mn@doi [\mnras] {10.1093/mnras/200.4.1067}, \href {https://ui.adsabs.harvard.edu/abs/1982MNRAS.200.1067O} {200, 1067}

\bibitem[\protect\citeauthoryear{{Pacholczyk}}{{Pacholczyk}}{1970}]{Pacholczyk70}
{Pacholczyk} A.~G.,  1970, {Radio astrophysics}.

\bibitem[\protect\citeauthoryear{{Padovani} \& {Giommi}}{{Padovani} \& {Giommi}}{1995}]{Padovani1995}
{Padovani} P.,  {Giommi} P.,  1995, \apj, \href {https://ui.adsabs.harvard.edu/abs/1995ApJ...444..567P} {444, 567}

\bibitem[\protect\citeauthoryear{{Paiano}, {Franceschini}  \& {Stamerra}}{{Paiano} et~al.}{2017a}]{Paiano2017A}
{Paiano} S.,  {Franceschini} A.,   {Stamerra} A.,  2017a, \mn@doi [\mnras] {10.1093/mnras/stx749}, \href {https://ui.adsabs.harvard.edu/abs/2017MNRAS.468.4902P} {468, 4902}

\bibitem[\protect\citeauthoryear{Paiano, Landoni, Falomo, Treves, Scarpa  \& Righi}{Paiano et~al.}{2017b}]{Paiano17}
Paiano S.,  Landoni M.,  Falomo R.,  Treves A.,  Scarpa R.,   Righi C.,  2017b, The Astrophysical Journal, 837, 144

\bibitem[\protect\citeauthoryear{{Pajdosz-{\'S}mierciak}, {{\'S}mierciak}  \& {Jamrozy}}{{Pajdosz-{\'S}mierciak} et~al.}{2022}]{Smierciak2022}
{Pajdosz-{\'S}mierciak} U.,  {{\'S}mierciak} B.,   {Jamrozy} M.,  2022, \mn@doi [\mnras] {10.1093/mnras/stac1372}, \href {https://ui.adsabs.harvard.edu/abs/2022MNRAS.514.2122P} {514, 2122}

\bibitem[\protect\citeauthoryear{{Peacock}}{{Peacock}}{1999}]{Peacock1999}
{Peacock} J.~A.,  1999, {Cosmological Physics}.
{Cambridge University Press}

\bibitem[\protect\citeauthoryear{{Perley} \& {Butler}}{{Perley} \& {Butler}}{2017}]{Perley2017}
{Perley} R.~A.,  {Butler} B.~J.,  2017, \mn@doi [\apjs] {10.3847/1538-4365/aa6df9}, \href {https://ui.adsabs.harvard.edu/abs/2017ApJS..230....7P} {230, 7}

\bibitem[\protect\citeauthoryear{{Perlman} \& {Stocke}}{{Perlman} \& {Stocke}}{1994}]{Perlman1994}
{Perlman} E.~S.,  {Stocke} J.~T.,  1994, \aj, 108, 56

\bibitem[\protect\citeauthoryear{{Petrov}, {Kovalev}  \& {Plavin}}{{Petrov} et~al.}{2019}]{Petrov2019}
{Petrov} L.,  {Kovalev} Y.~Y.,   {Plavin} A.~V.,  2019, \mn@doi [\mnras] {10.1093/mnras/sty2807}, \href {https://ui.adsabs.harvard.edu/abs/2019MNRAS.482.3023P} {482, 3023}

\bibitem[\protect\citeauthoryear{{Piner}, {Pant}  \& {Edwards}}{{Piner} et~al.}{2010}]{Piner2010}
{Piner} B.~G.,  {Pant} N.,   {Edwards} P.~G.,  2010, \mn@doi [\apj] {10.1088/0004-637X/723/2/1150}, \href {https://ui.adsabs.harvard.edu/abs/2010ApJ...723.1150P} {723, 1150}

\bibitem[\protect\citeauthoryear{{Piranomonte}, {Perri}, {Giommi}, {Landt}  \& {Padovani}}{{Piranomonte} et~al.}{2007}]{Piranomonte2007}
{Piranomonte} S.,  {Perri} M.,  {Giommi} P.,  {Landt} H.,   {Padovani} P.,  2007, \mn@doi [\aap] {10.1051/0004-6361:20077086}, \href {https://ui.adsabs.harvard.edu/abs/2007A&A...470..787P} {470, 787}

\bibitem[\protect\citeauthoryear{{Punch} et~al.,}{{Punch} et~al.}{1992}]{Punch1992}
{Punch} M.,  et~al., 1992, \mn@doi [\nat] {10.1038/358477a0}, \href {https://ui.adsabs.harvard.edu/abs/1992Natur.358..477P} {358, 477}

\bibitem[\protect\citeauthoryear{{Punsly}, {Tramacere}, {Kharb}  \& {Marziani}}{{Punsly} et~al.}{2018}]{Punsly2018}
{Punsly} B.,  {Tramacere} A.,  {Kharb} P.,   {Marziani} P.,  2018, \mn@doi [\apj] {10.3847/1538-4357/aaefe7}, \href {https://ui.adsabs.harvard.edu/abs/2018ApJ...869..174P} {869, 174}

\bibitem[\protect\citeauthoryear{{Pushkarev}, {Gabuzda}, {Vetukhnovskaya}  \& {Yakimov}}{{Pushkarev} et~al.}{2005}]{Pushkarev2005}
{Pushkarev} A.~B.,  {Gabuzda} D.~C.,  {Vetukhnovskaya} Y.~N.,   {Yakimov} V.~E.,  2005, \mn@doi [\mnras] {10.1111/j.1365-2966.2004.08535.x}, \href {https://ui.adsabs.harvard.edu/abs/2005MNRAS.356..859P} {356, 859}

\bibitem[\protect\citeauthoryear{{Pushkarev}, {Kovalev}, {Lister}, {Savolainen}, {Aller}, {Aller}  \& {Hodge}}{{Pushkarev} et~al.}{2017}]{Pushkarev2017}
{Pushkarev} A.,  {Kovalev} Y.,  {Lister} M.,  {Savolainen} T.,  {Aller} M.,  {Aller} H.,   {Hodge} M.,  2017, \mn@doi [Galaxies] {10.3390/galaxies5040093}, \href {https://ui.adsabs.harvard.edu/abs/2017Galax...5...93P} {5, 93}

\bibitem[\protect\citeauthoryear{{Pushkarev} et~al.,}{{Pushkarev} et~al.}{2023}]{Pushkarev2023}
{Pushkarev} A.~B.,  et~al., 2023, \mn@doi [\mnras] {10.1093/mnras/stad525}, \href {https://ui.adsabs.harvard.edu/abs/2023MNRAS.520.6053P} {520, 6053}

\bibitem[\protect\citeauthoryear{{Rau} \& {Cornwell}}{{Rau} \& {Cornwell}}{2011}]{MTMFS2011}
{Rau} U.,  {Cornwell} T.~J.,  2011, \mn@doi [\aap] {10.1051/0004-6361/201117104}, \href {https://ui.adsabs.harvard.edu/abs/2011A&A...532A..71R} {532, A71}

\bibitem[\protect\citeauthoryear{{Rector}, {Gabuzda}  \& {Stocke}}{{Rector} et~al.}{2003}]{Rector2003}
{Rector} T.~A.,  {Gabuzda} D.~C.,   {Stocke} J.~T.,  2003, \mn@doi [\aj] {10.1086/367802}, \href {https://ui.adsabs.harvard.edu/abs/2003AJ....125.1060R} {125, 1060}

\bibitem[\protect\citeauthoryear{{Rovero}, {Muriel}, {Donzelli}  \& {Pichel}}{{Rovero} et~al.}{2016}]{Rovero2016}
{Rovero} A.~C.,  {Muriel} H.,  {Donzelli} C.,   {Pichel} A.,  2016, \mn@doi [\aap] {10.1051/0004-6361/201527778}, \href {https://ui.adsabs.harvard.edu/abs/2016A&A...589A..92R} {589, A92}

\bibitem[\protect\citeauthoryear{{Saikia} \& {Salter}}{{Saikia} \& {Salter}}{1988}]{Saikia1988}
{Saikia} D.~J.,  {Salter} C.~J.,  1988, \mn@doi [\araa] {10.1146/annurev.aa.26.090188.000521}, \href {https://ui.adsabs.harvard.edu/abs/1988ARA&A..26...93S} {26, 93}

\bibitem[\protect\citeauthoryear{Schmidt \& Green}{Schmidt \& Green}{1983}]{Schmidt1983}
Schmidt M.,  Green R.,  1983, \apj, 269, 352

\bibitem[\protect\citeauthoryear{{Seielstad}, {Pearson}  \& {Readhead}}{{Seielstad} et~al.}{1983}]{Seielstad1983}
{Seielstad} G.~A.,  {Pearson} T.~J.,   {Readhead} A.~C.~S.,  1983, \mn@doi [\pasp] {10.1086/131261}, \href {https://ui.adsabs.harvard.edu/abs/1983PASP...95..842S} {95, 842}

\bibitem[\protect\citeauthoryear{{Shi}, {Liu}  \& {Song}}{{Shi} et~al.}{2007}]{Shi2007}
{Shi} W.,  {Liu} X.,   {Song} H.,  2007, \mn@doi [\apss] {10.1007/s10509-007-9413-z}, \href {https://ui.adsabs.harvard.edu/abs/2007Ap&SS.310...59S} {310, 59}

\bibitem[\protect\citeauthoryear{{Sillanpaa}, {Haarala}, {Valtonen}, {Sundelius}  \& {Byrd}}{{Sillanpaa} et~al.}{1988}]{Sillanpaa1988}
{Sillanpaa} A.,  {Haarala} S.,  {Valtonen} M.~J.,  {Sundelius} B.,   {Byrd} G.~G.,  1988, \mn@doi [\apj] {10.1086/166033}, \href {https://ui.adsabs.harvard.edu/abs/1988ApJ...325..628S} {325, 628}

\bibitem[\protect\citeauthoryear{{Sillanpaa} et~al.,}{{Sillanpaa} et~al.}{1996}]{Sillanpaa1996}
{Sillanpaa} A.,  et~al., 1996, \aap, \href {https://ui.adsabs.harvard.edu/abs/1996A&A...305L..17S} {305, L17}

\bibitem[\protect\citeauthoryear{Silpa, {Kharb}, {Harrison}, {Ho}, {Jarvis}, {Ishwara-Chandra}  \& {Sebastian}}{Silpa et~al.}{2021}]{Silpa2021}
Silpa S.,  {Kharb} P.,  {Harrison} C.~M.,  {Ho} L.~C.,  {Jarvis} M.~E.,  {Ishwara-Chandra} C.~H.,   {Sebastian} B.,  2021, \mnras, \href {https://ui.adsabs.harvard.edu/abs/2021MNRAS.507..991S} {507, 991}

\bibitem[\protect\citeauthoryear{{Stickel}, {Padovani}, {Urry}, {Fried}  \& {Kuehr}}{{Stickel} et~al.}{1991}]{Stickel1991}
{Stickel} M.,  {Padovani} P.,  {Urry} C.~M.,  {Fried} J.~W.,   {Kuehr} H.,  1991, \apj, \href {https://ui.adsabs.harvard.edu/abs/1991ApJ...374..431S} {374, 431}

\bibitem[\protect\citeauthoryear{{Stocke}, {Morris}, {Gioia}, {Maccacaro}, {Schild}, {Wolter}, {Fleming}  \& {Henry}}{{Stocke} et~al.}{1991}]{Stocke1991}
{Stocke} J.~T.,  {Morris} S.~L.,  {Gioia} I.~M.,  {Maccacaro} T.,  {Schild} R.,  {Wolter} A.,  {Fleming} T.~A.,   {Henry} J.~P.,  1991, \apjs, \href {https://ui.adsabs.harvard.edu/abs/1991ApJS...76..813S} {76, 813}

\bibitem[\protect\citeauthoryear{{Taylor}, {Stil}  \& {Sunstrum}}{{Taylor} et~al.}{2009}]{Taylor2009}
{Taylor} A.~R.,  {Stil} J.~M.,   {Sunstrum} C.,  2009, \mn@doi [\apj] {10.1088/0004-637X/702/2/1230}, \href {https://ui.adsabs.harvard.edu/abs/2009ApJ...702.1230T} {702, 1230}

\bibitem[\protect\citeauthoryear{{Urry} \& {Padovani}}{{Urry} \& {Padovani}}{1995}]{Urry95}
{Urry} C.~M.,  {Padovani} P.,  1995, \pasp, \href {https://ui.adsabs.harvard.edu/abs/1995PASP..107..803U} {107, 803}

\bibitem[\protect\citeauthoryear{{Valtonen} et~al.,}{{Valtonen} et~al.}{2006}]{Valtonen2006}
{Valtonen} M.~J.,  et~al., 2006, \mn@doi [\apjl] {10.1086/505039}, \href {https://ui.adsabs.harvard.edu/abs/2006ApJ...643L...9V} {643, L9}

\bibitem[\protect\citeauthoryear{{Vigotti}, {Grueff}, {Perley}, {Clark}  \& {Bridle}}{{Vigotti} et~al.}{1989}]{Vigotti1989}
{Vigotti} M.,  {Grueff} G.,  {Perley} R.,  {Clark} B.~G.,   {Bridle} A.~H.,  1989, \mn@doi [\aj] {10.1086/115153}, \href {https://ui.adsabs.harvard.edu/abs/1989AJ.....98..419V} {98, 419}

\bibitem[\protect\citeauthoryear{Wardle}{Wardle}{2018}]{Wardle2018}
Wardle J.,  2018, Galaxies, 6, 5

\bibitem[\protect\citeauthoryear{{Willis}, {Strom}, {Bridle}  \& {Fomalont}}{{Willis} et~al.}{1981}]{Willis1981}
{Willis} A.~G.,  {Strom} R.~G.,  {Bridle} A.~H.,   {Fomalont} E.~B.,  1981, \aap, \href {https://ui.adsabs.harvard.edu/abs/1981A&A....95..250W} {95, 250}

\bibitem[\protect\citeauthoryear{{Wilson}, {Ward}, {Axon}, {Elvis}  \& {Meurs}}{{Wilson} et~al.}{1979}]{Wilson1979}
{Wilson} A.~S.,  {Ward} M.~J.,  {Axon} D.~J.,  {Elvis} M.,   {Meurs} E.~J.~A.,  1979, \mn@doi [\mnras] {10.1093/mnras/187.2.109}, \href {https://ui.adsabs.harvard.edu/abs/1979MNRAS.187..109W} {187, 109}

\bibitem[\protect\citeauthoryear{{Wu}, {Jiang}, {Gu}  \& {Liu}}{{Wu} et~al.}{2007}]{Wu2007}
{Wu} Z.,  {Jiang} D.~R.,  {Gu} M.,   {Liu} Y.,  2007, \mn@doi [\aap] {10.1051/0004-6361:20066754}, \href {https://ui.adsabs.harvard.edu/abs/2007A&A...466...63W} {466, 63}

\bibitem[\protect\citeauthoryear{Wu, Gu  \& Jiang}{Wu et~al.}{2009}]{Wu_2009}
Wu Z.-Z.,  Gu M.-F.,   Jiang D.-R.,  2009, \mn@doi [Res. \aap] {10.1088/1674-4527/9/2/006}, 9, 168

\bibitem[\protect\citeauthoryear{{Xu}, {Readhead}, {Pearson}, {Polatidis}  \& {Wilkinson}}{{Xu} et~al.}{1995}]{Xu1995}
{Xu} W.,  {Readhead} A.~C.~S.,  {Pearson} T.~J.,  {Polatidis} A.~G.,   {Wilkinson} P.~N.,  1995, \mn@doi [\apjs] {10.1086/192189}, \href {https://ui.adsabs.harvard.edu/abs/1995ApJS...99..297X} {99, 297}

\bibitem[\protect\citeauthoryear{{Zavala} \& {Taylor}}{{Zavala} \& {Taylor}}{2004}]{Zavala2004}
{Zavala} R.~T.,  {Taylor} G.~B.,  2004, \mn@doi [\apj] {10.1086/422741}, \href {https://ui.adsabs.harvard.edu/abs/2004ApJ...612..749Z} {612, 749}

\bibitem[\protect\citeauthoryear{{Zhang} \& {Baath}}{{Zhang} \& {Baath}}{1990}]{Zhang1990}
{Zhang} F.~J.,  {Baath} L.~B.,  1990, \aap, \href {https://ui.adsabs.harvard.edu/abs/1990A&A...236...47Z} {236, 47}

\bibitem[\protect\citeauthoryear{{Zhang} \& {Baath}}{{Zhang} \& {Baath}}{1991}]{Zhang1991}
{Zhang} F.~J.,  {Baath} L.~B.,  1991, \mn@doi [\mnras] {10.1093/mnras/248.3.566}, \href {https://ui.adsabs.harvard.edu/abs/1991MNRAS.248..566Z} {248, 566}

\bibitem[\protect\citeauthoryear{{de Bruyn} \& {Schilizzi}}{{de Bruyn} \& {Schilizzi}}{1986}]{deBruyn1986}
{de Bruyn} A.~G.,  {Schilizzi} R.~T.,  1986, in {Swarup} G.,  {Kapahi} V.~K.,  eds, ~ Vol. 119, Quasars. p.~203

\bibitem[\protect\citeauthoryear{{van Moorsel}, {Kemball}  \& {Greisen}}{{van Moorsel} et~al.}{1996}]{vanMoorsel1996}
{van Moorsel} G.,  {Kemball} A.,   {Greisen} E.,  1996, in {Jacoby} G.~H.,  {Barnes} J.,  eds,  Astronomical Society of the Pacific Conference Series Vol. 101, Astronomical Data Analysis Software and Systems V. p.~37

\makeatother
\end{thebibliography}
\end{document}